\documentclass[english]{vc}

\title{Neural Scene Baking for Permutation Invariant Transparency Rendering with Real-time Global Illumination}

\plainauthors{Ziyang Zhang, Edgar Simo-Serra}

\begin{authors}
  \newcolumntype{C}{>{\centering}p{0.20\textwidth}}

  \begin{tabular}{CC}
    Ziyang Zhang$^\dagger$ & Edgar Simo-Serra$^\ddagger$ 
  \end{tabular}
\end{authors}

\begin{affiliations}
  \begin{tabular}{cc}
    Waseda University
  \end{tabular}
\end{affiliations}

\begin{emails}
  \begin{tabular}{c}
    E-mail: $\dagger${}ziyangz5@toki.waseda.jp, $\ddagger${}ess@waseda.jp
  \end{tabular}
\end{emails}

\newcommand{\onedot}{.}

\newcommand{\ie}{\emph{i.e}\onedot}

\newcommand{\etal}{\emph{et al}\onedot}

\newcommand{\hide}[1]{}

\usepackage{adjustbox}
\usepackage{multicol}
\usepackage{microtype}
\usepackage{booktabs}
\usepackage{amsmath}
\usepackage{multirow}
\usepackage{subcaption}
\usepackage{pgfplots}
\usepackage{float}

\begin{abstract}
  Neural rendering provides a fundamentally new way to render photorealistic images. Similar to traditional light-baking methods, neural rendering utilizes neural networks to bake representations of scenes, materials, and lights into latent vectors learned from path-tracing ground truths. However, existing neural rendering algorithms typically use G-buffers to provide position, normal, and texture information of scenes, which are prone to occlusion by transparent surfaces, leading to distortions and loss of detail in the rendered images. To address this limitation, we propose a novel neural rendering pipeline that accurately renders the scene behind transparent surfaces with global illumination and variable scenes. Our method separates the G-buffers of opaque and transparent objects, retaining G-buffer information behind transparent objects. Additionally, to render the transparent objects with permutation invariance, we designed a new permutation-invariant neural blending function. We integrate our algorithm into an efficient custom renderer to achieve real-time performance. Our results show that our method is capable of rendering photorealistic images with variable scenes and viewpoints, accurately capturing complex transparent structures along with global illumination. Our renderer can achieve real-time performance ($256\times 256$ at 63 FPS and $512\times 512$ at 32 FPS) on scenes with multiple variable transparent objects.
\end{abstract}

\begin{teaserfigure}
  \centering
  \includegraphics[width=\textwidth]{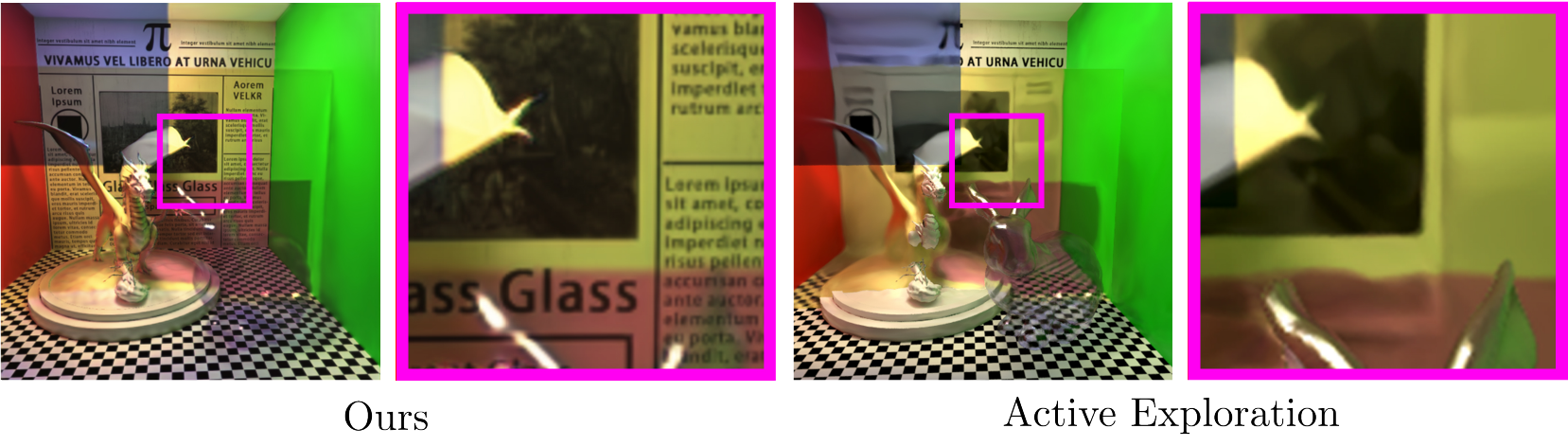}
  \vspace*{-10mm}
  \caption{\textbf{Our primary results.} Our custom neural rendering framework addresses transparency 
  issues in existing neural rendering approaches. The loss of details of our method is significant lower than the Active Exploration\cite{diolatzis2022active}.}
  \label{fig:teaser}
\end{teaserfigure}

\begin{document}

\maketitle

\section{Introduction}
\begin{figure*}[h]
    \includegraphics[width=\textwidth]{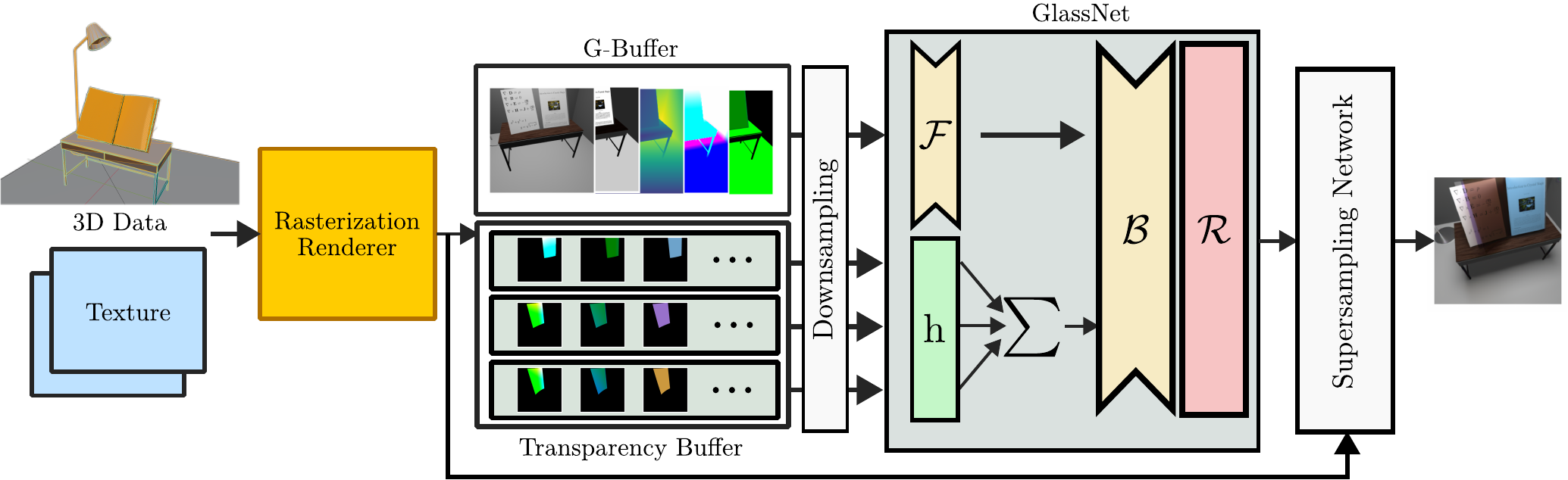}
    \caption{\textbf{Overview of our neural scene rendering framework.}
    First, our rasterization-based renderer will first render the
    G-buffer, direct lighting, and transparency buffers. Then, a neural network,
    which we denote as \textit{GlassNet}, will use those buffers and rendering
    results as inputs to synthesize high quality images with global illumination
    and accurate transparency. Details of \textit{GlassNet} can be found
    in \cref{sec:NeuralRenderer} and \cref{fig:net_structure}. }
    \label{fig:RPStructure}
\end{figure*}

\begin{figure*}[h]
    \includegraphics[width=\textwidth]{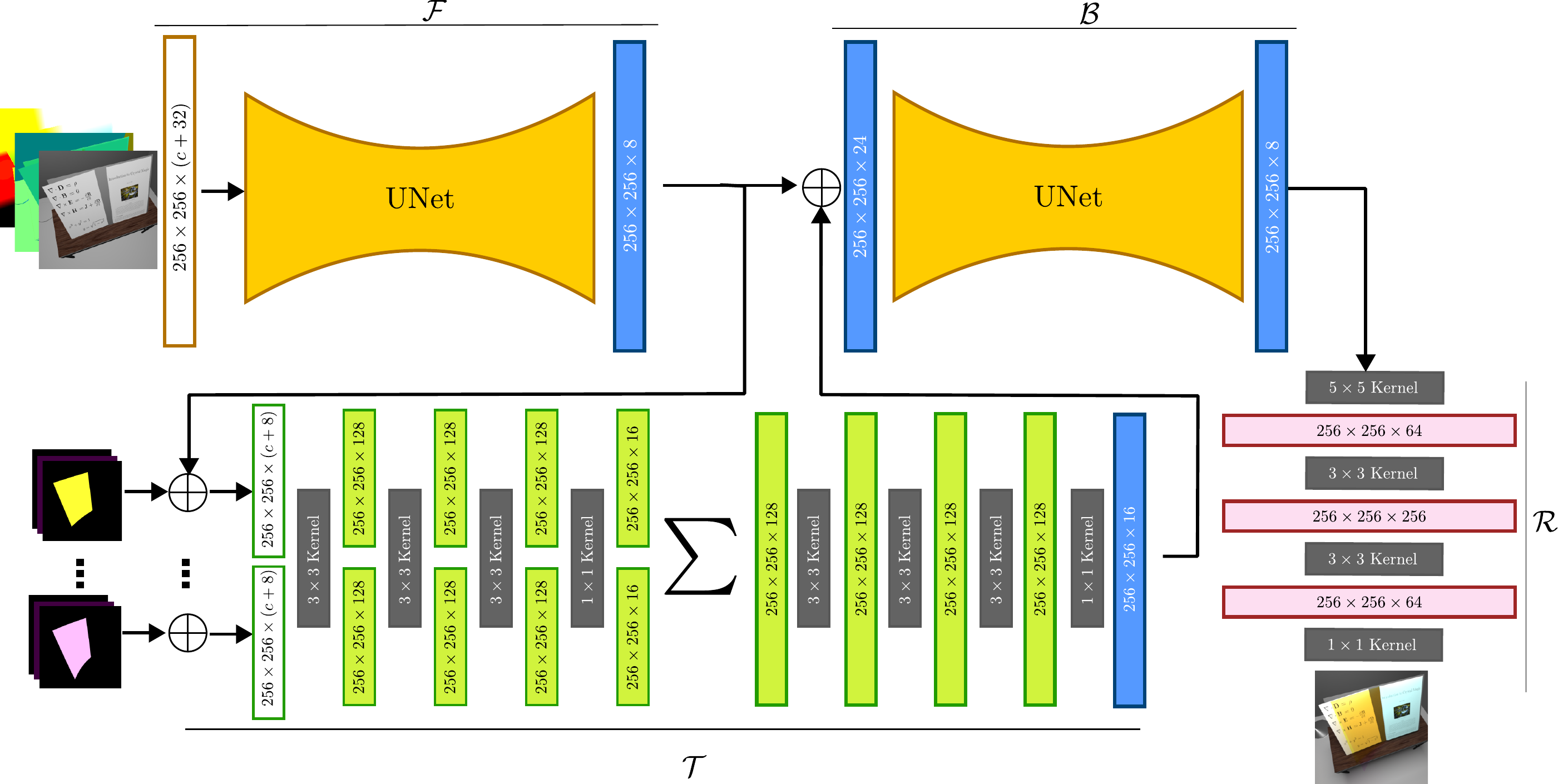}
    \caption{\textbf{Architecture of GlassNet.} Our proposed method, GlassNet, contains four building blocks including the scene encoder, $\mathcal{F}$; the permutation
    invariant transparency buffer blending function, $\mathcal{T}$; the final
    blending network, $\mathcal{B}$; and the rendering network, $\mathcal{R}$. All parts are trained jointly.}
    
    \label{fig:net_structure}
  \end{figure*}
Global illumination rendering has been a fundamental challenge in computer
graphics due to the complexity of the physical phenomena involved.
Traditionally, path tracing
algorithms~\cite{kajiya1986rendering,lafortune1996rendering} have been the main
method to render photorealistic images with high-quality global illumination
effects. However, such methods are known to be expensive and time-consuming;
rendering one image may take minutes. Baked global illumination techniques, on
the other hand, provide efficient solutions for interactive rendering by
pre-computing lighting information. Methods such as lightmap
baking and light probes~\cite{greger1998irradiance} have
been commonly used in game engines to generate real-time global illumination, but they have limitations on dynamic scenes.

Recently, neural rendering approaches have emerged as promising alternative for
realistic image synthesis. Using neural networks to represent scenes,
materials, and lighting can be seen as one type of pre-computation method. In the
early stage, neural rendering methods were only capable of rendering fixed
scenes under limited types of lighting with complex spatial
inputs~\cite{ren2013global}. More recently, by using G-buffer as inputs, newer
methods avoid using complex spatial data such as point clouds, and are able to
render scenes with variable geometries, materials, and
lighting~\cite{eslami2018neural,granskog2020compositional,rainer2022neural,diolatzis2022active,gao2022neural}.
However, the usages of G-buffer re-introduces the classic problem of
transparency rendering. When generating G-buffers, the shading information
will be overwritten by transparent surfaces in the front, as shown in \cref{fig:GBuffer}. Without such crucial shading
information, rendering models can only hard code all the information into the
neural network based on ground truths, which is unrealistic and
ineffective, leading to poor quality.

To address this limitation, we propose a novel neural rendering framework that
separately renders opaque objects and semi-transparent surfaces into two groups
of G-buffers. By doing this, all necessary shading information of both opaque
and transparent surfaces are preserved and can be used for further rendering.
Furthermore, we design a neural blending function to encode the transparency
buffer into a latent representation with permutation invariance to achieve
Order-Independent-Transparency (OIT). Finally, we use a deep CNN (Convolutional
Neural Network) to encode scenes and blend them with transparency
representations, rendering the resulting image with global illumination.

Our results show that we significantly improve the transparency quality,
exceeding the most recent neural approaches, Active
Exploration~\cite{diolatzis2022active} and Neural Global
Illumination~\cite{gao2022neural}.
Our method can accurately render the geometry and texture details under
transparent surfaces, whereas methods without transparency buffers converge to
low-quality results.

In summary, our main contributions are:
\begin{enumerate}
    \itemsep0em
    \item A novel neural rendering framework preserves both opaque and
    transparent information simultaneously on buffers, providing accurate
    shading information on rendering transparency.
    \item A neural blending function that encodes transparent surfaces
    with permutation invariant, achieving OIT.
    \item A real-time renderer capable of rendering images with high performance enabled by the above algorithms.
\end{enumerate}

\section{Related Work}
\subsection{Traditional Transparency Rendering}
In the traditional forward rendering pipeline, various techniques have been used to
handle transparency. One widely used algorithm is alpha
blending~\cite{porter1984compositing}, which utilizes a separate alpha channel
to control the color composition of semi-transparent objects with opaque objects
on each fragment. However, alpha blending is order-dependent and requires
sorting of objects by depth, which can be impractical. The high
computational complexity is not the only problem of those sorting-based
algorithms. When transparent objects intersect with each other, it is
impossible to sort the objects globally unless the sorting is done by
fragments, which again greatly increases the computational complexity. Another
algorithm called the A-Buffer algorithm~\cite{carpenter1984buffer} stores a
list of relevant fragments in each pixel, but sorting of all fragments is
required with unbounded memory usage. The $Z^3$ algorithm~\cite{jouppi1999z} is
a similar algorithm with bounded memory usage but comes with a trade-off in
image quality. In contrast, our proposed method does not require any
sorting, and does not generate a large amount of frame buffers during rendering
procedure.

To avoid sorting, several techniques have been introduced. Depth
peeling~\cite{everitt2001interactive} renders objects in multiple passes,
peeling transparent objects based on the z-buffer. However, the number of
passes required by depth peeling increases the computational cost.

These algorithms often encounter challenges in the modern deferred rendering
pipeline. Deferred rendering~\cite{saito1990comprehensible} is a technique that
separates the geometric stage and lighting stage into two passes. In the
geometry pass, all the geometry information is rendered into G-buffers to be
used by the lighting pass for shading. While deferred rendering significantly
improves rendering performance with multiple light sources, it introduces
complexity to transparency rendering. Rendering geometry information into the
G-buffer causes the objects behind transparent surfaces to be occluded, making
it impossible to retain all geometry information without a prohibitively large
number of buffers. Several algorithms solve this problem with worsened image
quality.~\cite{pangerl2009deferred, mara2013lighting}. By applying a
permutation-invariant blending algorithm, our method can combine generated
transparency information during inference, while only requiring two frame buffers
regardless of the number of transparent objects, significantly lowering the
memory usage.

Several neural rendering approaches utilize the G-buffers as input for their
neural
networks~\cite{granskog2020compositional,diolatzis2022active,gao2022neural, xin2020lightweight},
resulting in these methods also encountering transparency occlusion issues with
the G-buffers, causing bad image quality. 
In contrast, our method is not affected by such issues and supports OIT rendering.

\subsection{Baked Global Illumination}
Different from expensive path tracing methods, which are the main approach to
render photorealistic images in the modern film industry~\cite{keller2015path}, baked
or precomputed global illumination methods are more often used in interactive
or real-time settings~\cite{ritschel2012state}.

Lightmap baking is a popular technique used widely in modern game
engines~\cite{o2018precomputed}. It precomputes the diffuse global
illumination and uses interpolation to generate global illumination during
rendering. Several methods have been proposed to generate the lightmap offline, such
as by radiosity~\cite{cohen1993radiosity} or
path-tracing. Naively, baked lightmap only supports
static scenes with diffuse objects. Light probes technique  was introduced to
enable global illumination for dynamic objects~\cite{greger1998irradiance}.
McGuire \etal~\cite{mcguire2017real} extended light probes to support glossy surfaces.
However, such methods are limited to either static lighting or a limited amount
of light transportation types.

Several neural approaches use neural networks to encode the scene
representation into neural vectors and synthesize global
illumination such as Compositional Neural
Scene Representations (CNSR)~\cite{granskog2020compositional}, Active Exploration and Neural Global Illumination.
Similar to such neural approaches, our method can render global illumination
with dynamic scenes with various light transportation types, with support for
transparent objects.

\subsection{Neural Rendering}\label{subsection:NeuralRendering}
Using neural networks for rendering is a rapidly developing research topic that
provides fundamentally new approaches to photorealistic rendering. We will
introduce the closest related paper to this work. We refer to~\cite{tewari2020state} for a more detailed neural rendering survey.

Initially, Ren \etal~\cite{ren2013global} used neural networks to learn
indirect lights from dynamic lighting and roughness with fixed geometry,
complex spatial data, and only point lights. Later, Eslami
\etal~\cite{eslami2018neural} proposed a method that trains an encoder-decoder
neural network to represent scenes as latent vectors. CNSR by Granskog \etal~further improved the interpretability
of the neural scene representation by partitioning the scene representation
vectors into sections of lighting, materials, geometry, etc.
Similar to our
method, most recent
solutions~\cite{granskog2020compositional,rainer2022neural,diolatzis2022active,gao2022neural}
use G-buffers as the input to the neural networks to produce indirect lighting
and global illumination. Using G-buffers instead of point clouds or voxel data simplifies the input data.
Such algorithms can usually be seen as precomputed light
transportation and a form of scene baking.
Using neural networks to precompute wavelet function has also shown strong improvement on glossy global illumination~\cite{raghavan2023neural}. Attempts on high performance real-time neural rendering are also made by Xin \etal~\cite{xin2020lightweight}, but only limited on single-bounce diffuse indirect illumination

However, using G-buffers as inputs brings back the classic issue caused by
transparent objects. The G-buffers cannot provide information on objects behind
transparent surfaces, ending with poor rendering quality by such methods. In
our framework, we use independent G-buffers for the transparent objects and a
permutation invariant neural network to achieve OIT. Note that the recent studies in neural rendering are mainly orthogonal
to our proposed method with different goals. Our method can be directly
combined with those methods to improve the rendering quality further.

\subsection{Permutation Invariant Machine Learning}
PointNet~\cite{qi2017pointnet} uses a multi-layer perceptron network to
transform the input into latent vectors and uses a symmetric function to
compose such vectors, which is proved to be a simple but strong method.
PointNet inspires how our rendering neural network can process each transparency
input to cancel the permutation order dependency. We refer a more detailed review of permutation-invariant neural network to Section 6 of \cite{bloem2020probabilistic}.

\section{Proposed Approach}

\subsection{Overview}

Given a scene, the goal of our approach is to bake a neural scene
representation into a neural network that can correctly
approximate global illumination with variable parameters of material, geometry,
and lighting without being affected by transparent surfaces. For any shading
position, $x$, on surfaces, the outgoing radiance, $L_o$, can be calculated by
the Rendering Equation\cite{kajiya1986rendering}:
\begin{equation}
\label{eq:RenderingEquation}
\begin{aligned}
    L_o(x,\omega_o)= &\int_\Omega L_i(x,\omega_i)f_x(x,\omega_o,\omega_i)|n_x\cdot\omega_i|\,\mathrm{d}\omega_i\\
    &+ L_e(x,\omega_o)
\end{aligned}
\end{equation}
\noindent where $f_x$ and $n_x$ represent the BSDF and normal; $\omega_o$
indicates outgoing or viewing direction, $\omega_i$ indicates the incident
direction; $L_i$ and $L_e$ denote the incident radiance and emission
radiance. $\Omega$ denotes a hemisphere.

The Rendering Equation can also be written into the following form such as the
direct lighting and the indirect lighting are separated:
\begin{equation}
\label{eq:RenderingEquation2}
    L_o(x,\omega_o)=L_e(x,\omega_o) + L_d(x,\omega_o) + L_g(x,\omega_i)
\end{equation}
\noindent where $L_d$ is the direct lighting, and $L_g$ is the indirect lighting which
contributes mainly to the global illumination. Current methods use
high-dimension complex non-linear functions, \ie, neural networks, to predict
$L_o(x,\omega_o)$ given all the necessary geometry, lighting, and material
parameters as inputs. In contrast, our approach consists of a neural network,
which we denote as \textit{GlassNet}, to predict the indirect lighting, $L_g$,
as shown in \cref{fig:RPStructure}. We use Linearly Transformed Cosines (LTCs) \cite{heitz2016real} as a sufficient approximation to the direct illumination term $L_d$ as
an input to our neural network to achieve efficiency. Notice that as an input to the neural network, $L_d$ is not necessarily to be analytically accurate because the neural network will learn the complexity of the lighting transportation from the path tracing ground truths.

\begin{figure}[t]
\centering
  \includegraphics[width=0.475\textwidth]{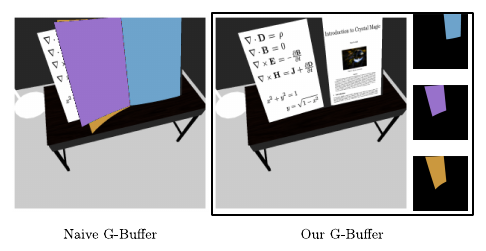}

  \caption{\textbf{Our G-buffer scene representation approach.}
  Instead of naively rendering the entire scene to a single set of G-buffers, we use separate buffers for the transparent objects, allowing us to represent complex transparency visibility in scenes accurately.}
  \label{fig:GBuffer}
\end{figure}

Most recent methods provide necessary shading information such as $\omega_o$,
position, texture, and material parameters through G-buffers. However, the naive approach using
G-buffers as input brings a significant problem: as shown in \cref{fig:GBuffer}, important shading information is hidden by
transparent surfaces.
We propose a new neural rendering framework that separately draws opaque
G-buffers and transparency buffers, preserving all the information, including
the transparent surfaces and the objects behind those.
Then, we introduce a symmetric neural function to cancel the permutation
dependency in the transparency buffers and generate a neural transparency
representations as shown in \cref{subsection:PITrans}.

\subsection{Transparency Buffers}\label{sec:TranspHandle}

\label{subsection:customRP}

\begin{figure*}[thbp]

\centering

\begin{tabular}{@{} c c c c c c c @{}}
  \setlength{\tabcolsep}{0.05pt}
  \begin{tabular}{@{} c c c c c c c c c @{}}
    \setlength{\tabcolsep}{0.05pt}
       & Normal  & Position  & Albedo & Roughness & $\omega_o$  & Depth & $L_d$ & \\[1ex]
       & \includegraphics[width=1.5cm,height=1.5cm]{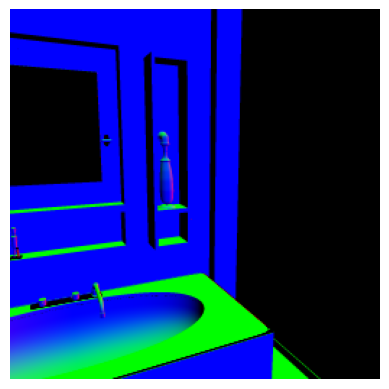} &
      \includegraphics[width=1.5cm,height=1.5cm]{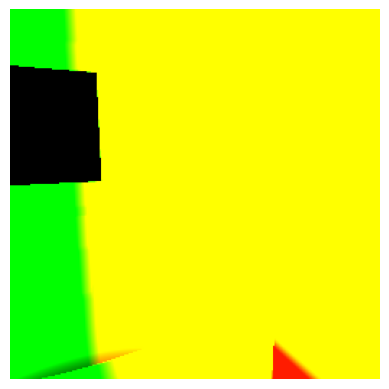} &
      \includegraphics[width=1.5cm,height=1.5cm]{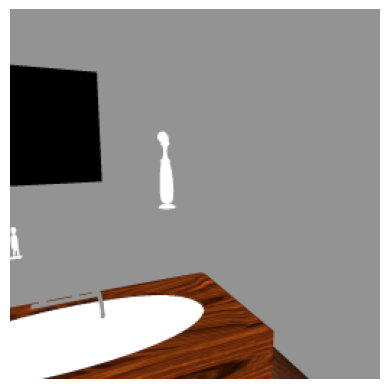} &
      \includegraphics[width=1.5cm,height=1.5cm]{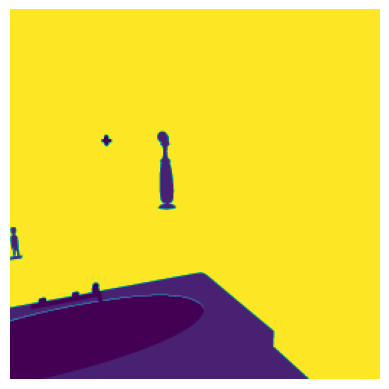} &
      \includegraphics[width=1.5cm,height=1.5cm]{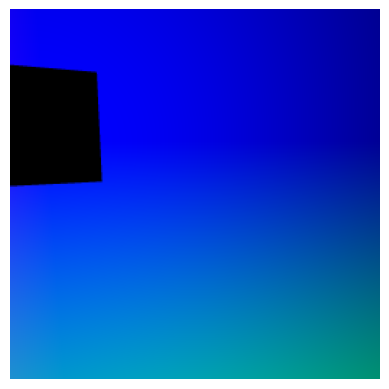} &
      \includegraphics[width=1.5cm,height=1.5cm]{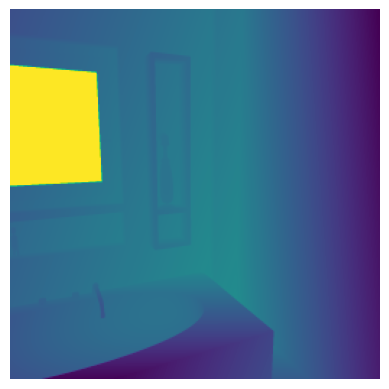} &
      \includegraphics[width=1.5cm,height=1.5cm]{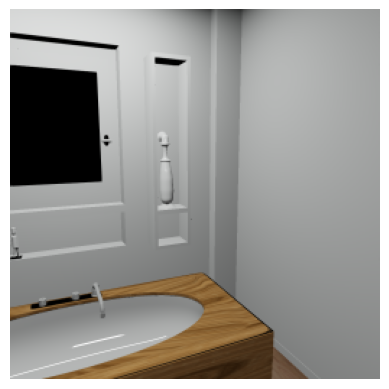} & \\[2ex]
      & \multicolumn{7}{c}{G-Buffer} \\[1ex]

      & Normal  & Position  & Albedo & Roughness & $\omega_o$  & Depth & Relative Depth & \\[1ex]
      & \includegraphics[width=1.5cm,height=1.5cm]{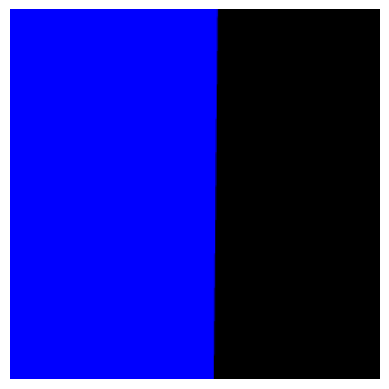} &
      \includegraphics[width=1.5cm,height=1.5cm]{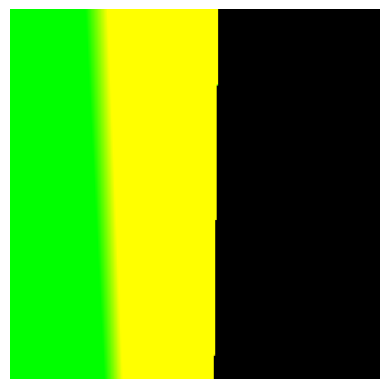} &
      \includegraphics[width=1.5cm,height=1.5cm]{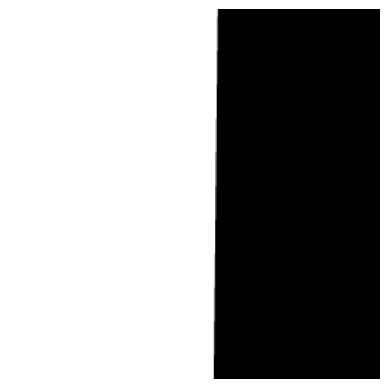} &
      \includegraphics[width=1.5cm,height=1.5cm]{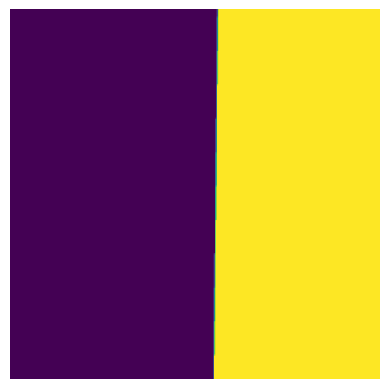} &
      \includegraphics[width=1.5cm,height=1.5cm]{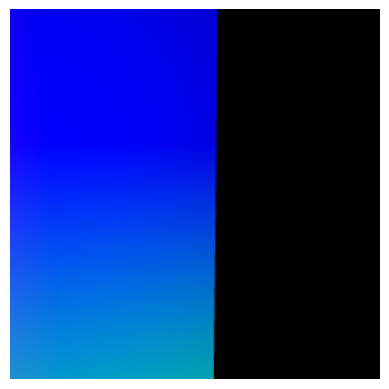} &
      \includegraphics[width=1.5cm,height=1.5cm]{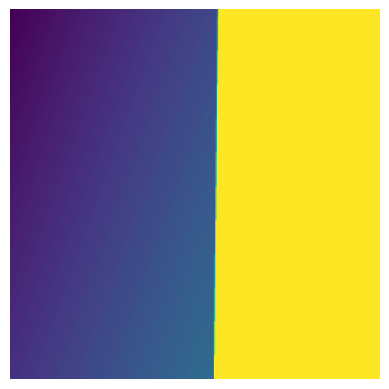} &
      \includegraphics[width=1.5cm,height=1.5cm]{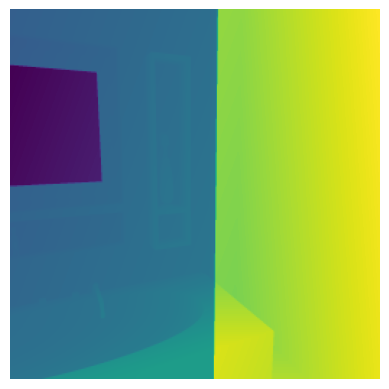} & \\[2ex]
      & \multicolumn{7}{c}{Transparency Buffer}
  \end{tabular}
&
  \begin{tabular}{@{} c @{}}
    Ground Truth \\[1ex]
    \includegraphics[width=3.5cm,height=3.5cm]{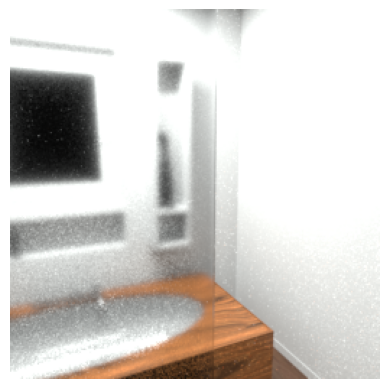}
  \end{tabular}
\end{tabular}

\caption{\textbf{G-buffer, transparency buffer, and ground truth.} Our
method can overcome the noise in the ground truth.}\label{whatever}
\label{fig:Inputs}
\end{figure*}

For a given scene, we rasterize each transparent surface into
a list of G-buffers containing all the information
required by a traditional path tracer to render $L_o$, including position,
surface normal, albedo, $\omega_o$, and material parameters as shown in \cref{fig:Inputs}. We generate a set of buffers with a size of
$(w, h, c*(t+1))$, where $w$, $h$ and $c$ denote the width, height and channel of images. $t$ is the number of transparent objects in the scene.
Finally, our rasterization-based renderer creates the opaque G-buffers, and use LTCs to compute the direct lighting $L_d$. By doing so, we can preserve all the necessary information on
both normal surfaces and transparent surfaces.

However, further processing is necessary for the naive transparency buffer to
be a good representation as a neural network input. Generally, neural networks
are known to depend on the order of inputs\cite{qi2017pointnet}.
Inputting the naive transparency buffer will cause our rendering network to be
order-dependent on transparency. With modern rendering engine optimization
features such as batch rendering\cite{wloka2003batch} and occlusion
culling\cite{pantazopoulos2002occlusion}, the order of rendered objects cannot
be guaranteed. The surfaces can even be culled during rendering. It is also
infeasible to sort the rendered buffers after rendering. In order to solve this
problem, we designed a symmetric neural network invariant to permutation to
blend the transparency buffer.

\subsection{Permutation Invariant Neural Representation}
\label{subsection:PITrans}
We design a blending invariant algorithm to cancel any sensitivity to the
order of the inputs while significantly lowering the memory usage during
inference as shown in \cref{sec:Performance}, inspired by traditional
transparency algorithms such as alpha blending. In alpha blending, the
function, $T$, of
compositing transparent objects can be described
as\cite{porter1984compositing,meshkin2007sort,salvi2014multi}:
\begin{equation}
   \label{eq:AlphaBlending}
   T\Bigl(\bigl\{(C_1,\alpha_1),(C_2,\alpha_2),\ldots,(C_t,\alpha_t)\bigr\}\Bigr) =
      \sum_{i=1}^t\alpha_i \, C_i \, z_i
\end{equation}
\noindent
where $C_i$ and $\alpha_i$ respectively are the color and the alpha
value of each transparent object. $z_i$ is the product of all alpha values in front of the object:
\begin{equation}
   \label{eq:AlphaBlendingZ}
    z_i = \begin{cases}
  1 & \text{if } i = t \\
  \prod_{j=1}^{t-i} (1-\alpha_j) & \text{otherwise}
\end{cases}
\end{equation}
\noindent
This function is clearly
input-order-dependent, but it can be revised into a function with
permutation invariant to inputs.

Based on the alpha blending function with the inspiration from the
PointNet, we design our permutation invariant transparency
buffer blending function, $\mathcal{T}$, to generate the neural
representation, $\tau$:

\begin{align}
   &\label{eq:SymmetricNN}
   \mathcal{T}\Bigl(\{b_1,b_2,\ldots,b_t\},\sigma\Bigr)\notag\\ &\approx g\Bigl(h(b_1,\sigma),h(b_2,\sigma),\ldots,h(b_t,\sigma)\Bigr) \noindent\notag\\
   &=\sum_{i=1}^th(b_i,\sigma)=\tau
\end{align}

\noindent where $b$ is the transparency buffer, $g$ is a symmetric function,
$\sigma$ is a scene representation, and $h$ is a neural network shared by all
the transparency buffers generating neural representations.
We use addition as $g$ to match the alpha blending function, which clearly makes this function mathematically permutation invariant to inputs. This
allows the neural network to blend all the transparency buffers into a single latent vector that is independent of the input order.

\subsection{Rendering Model}
\label{sec:NeuralRenderer}
We introduce our rendering model, \textit{GlassNet} as shown in \cref{fig:RPStructure}. \textit{GlassNet} has
four components: the scene encoder, $\mathcal{F}$; the permutation
invariant transparency buffer blending function, $\mathcal{T}$; the final
blending network, $\mathcal{B}$; and the rendering network, $\mathcal{R}$.
Both $\mathcal{F}$ and $\mathcal{B}$ use U-Net~\cite{isola2017image} as the
network backend, whereas other networks use multi-layer CNN without
down-sampling. All four components are trained jointly as one neural network. Full details of the model are shown in \cref{fig:net_structure}.

The scene encoding network, $\mathcal{F}$, accepts tensors
containing the position, normal, texture, material parameters, $\omega_o$,
depth buffer, and direct lighting. It then generates a neural scene representation,
$\sigma$. Similarly, each transparency buffer is processed by $\mathcal{T}$, generating the neural representation of transparent surfaces $\rho$. All the
inputs use positional encoding by Mildenhall
\etal~\cite{mildenhall2021nerf} to capture high-frequency information.
Afterward, the blending network $\mathcal{B}$ blends the scene representation $\sigma$ and transparent object representations $\rho$ and processes visibility. By doing so, $\mathcal{B}$ generates the final neural buffer, $\phi$.

Finally, the rendering network $\mathcal{R}$ takes the neural buffer $\phi$ with additional direct
lighting as input to render the final image. Although previous works, such as Neural Global Illumination, report using convectional neural network structure on the
rendering network can lead to distortion of shapes, we find that this problem
can usually be solved by using kernels with the size of $1\times1$ at the end
of the network, and adding larger kernel can help with global effects such as
blurring. For balancing the render quality and performance, the
\textit{GlassNet} is trained with lower resolution ($256\times256$) textures,
G-buffers, and ground truths. A multi-layer CNN is used as a super-sampling
network that super-samples the result at $\times2$ the original resolution.
Given that we can cheaply get high-resolution texture and G-buffer from the
rasterizer, the super-sampling network can be simple to achieve real-time
performance.

\subsection{Training}

We train the model in an end-to-end fashion to bake a specific scene by
randomly sampling camera angles and positions within certain ranges of each
scene following uniform distributions to render ground truth images.
In particular, for each scene, we generate output image and G-buffer
pairs for training.  The ground truth is rendered by the modern path tracer,
Mitsuba 3~\cite{Jakob2020DrJit}. For the G-buffer and direct lighting generation, we implemented our custom rasterization renderer using OpenGL in C++, with CUDA support to support neural network inference. As a loss function, we choose is the combination of
Structural Similarity Index Measure (SSIM)~\cite{wang2004image} and $L_1$ loss.

\section{Experiments}
\label{subsection:compare}
\newcommand{\compimage}[1]{\adjustbox{valign=m,vspace=0.25pt}{\includegraphics[width=.285\linewidth]{figures/comparison/#1}}}

\begin{figure}[t]
  \centering
    \includegraphics[width=0.475\textwidth]{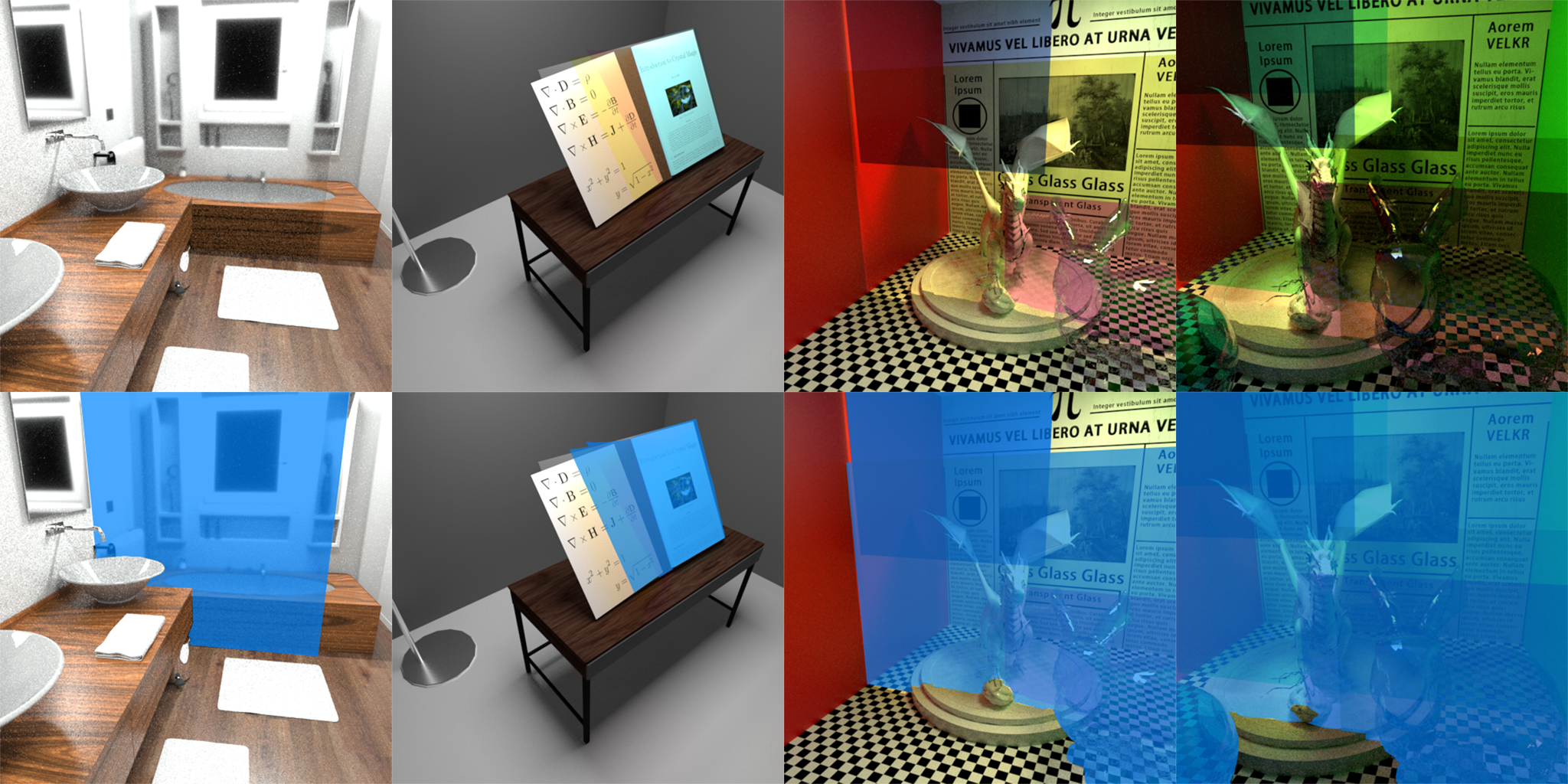}

    \caption{\textbf{Randomly selected images from datasets.}
    The transparent areas are highlighted by the blue masks in the second row.}
    \label{fig:ds}
\end{figure}

\begin{figure*}[ph!]
\centering
\begin{tabular}{cccc}
 & Ground Truth & Ours & Active Exploration \\
\rotatebox[origin=c]{90}{BATHROOM}  & \compimage{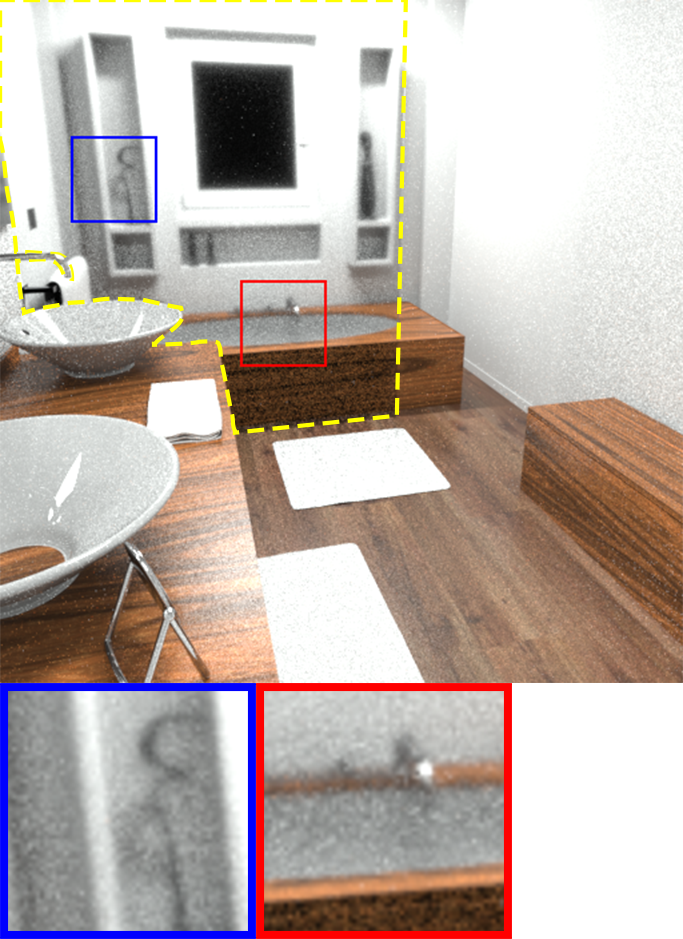}  & \compimage{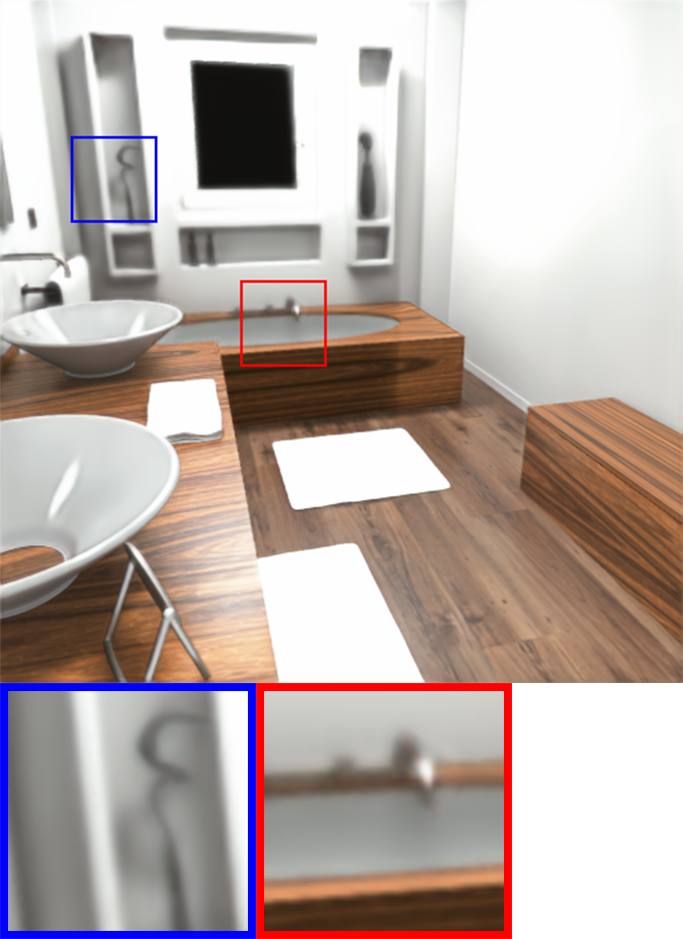} & \compimage{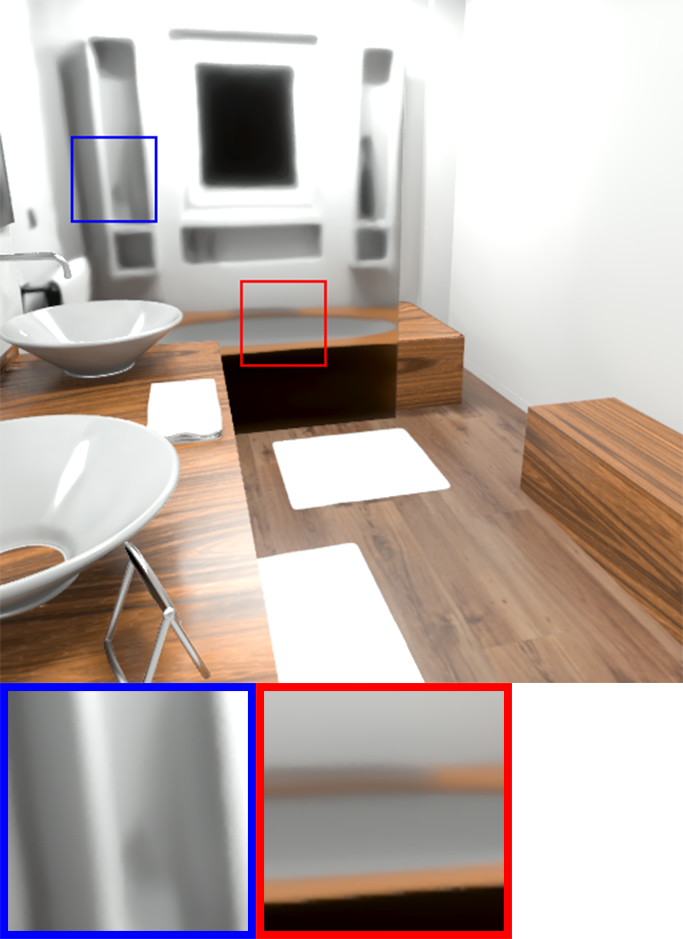} \\
\rotatebox[origin=c]{90}{DESK}  & \compimage{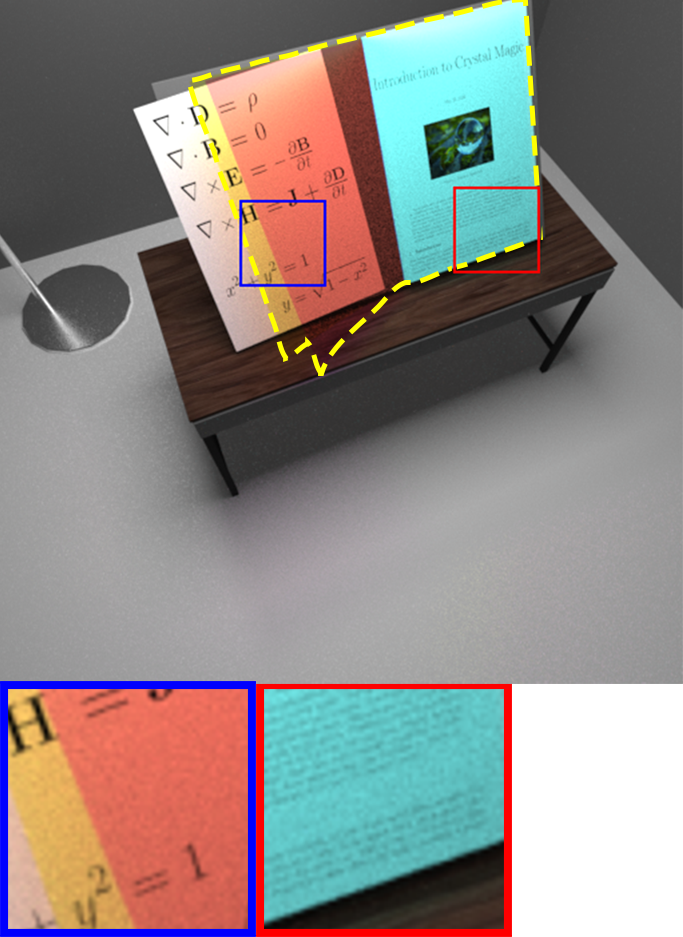}  & \compimage{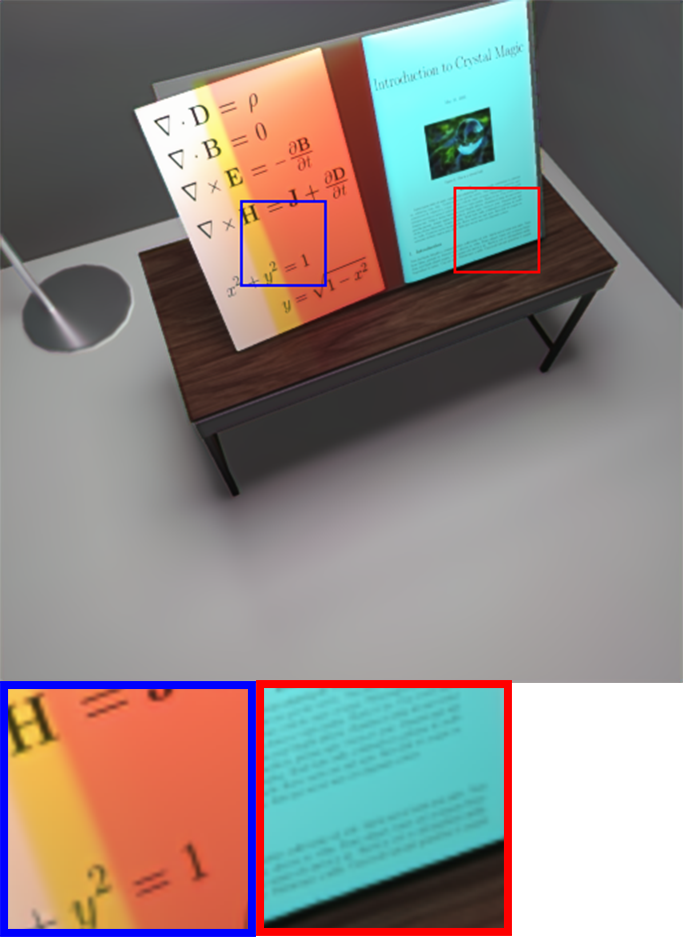} & \compimage{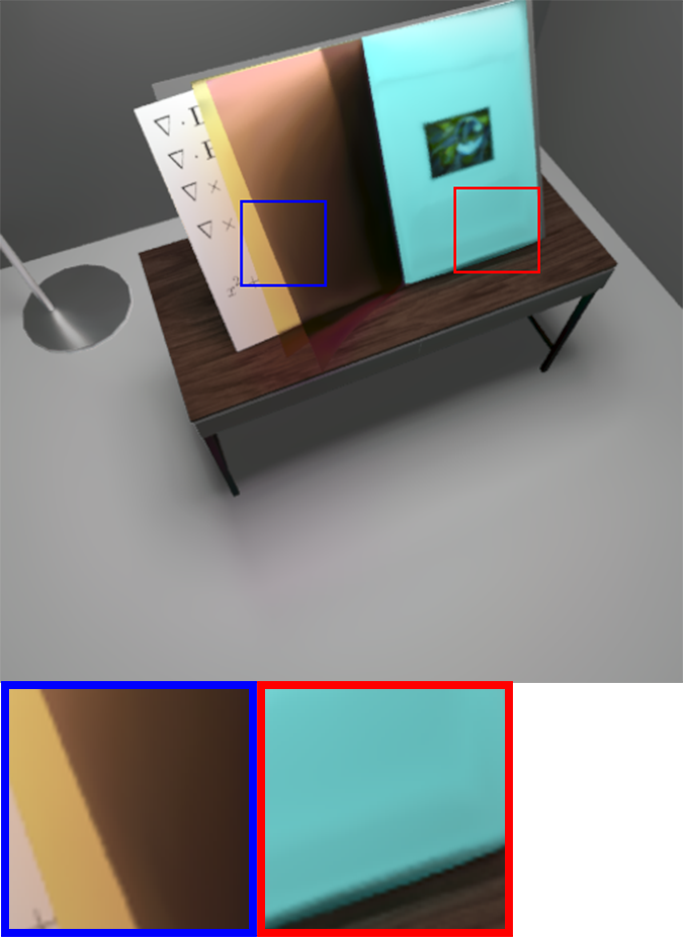} \\
\rotatebox[origin=c]{90}{CORNELLBOX}  & \compimage{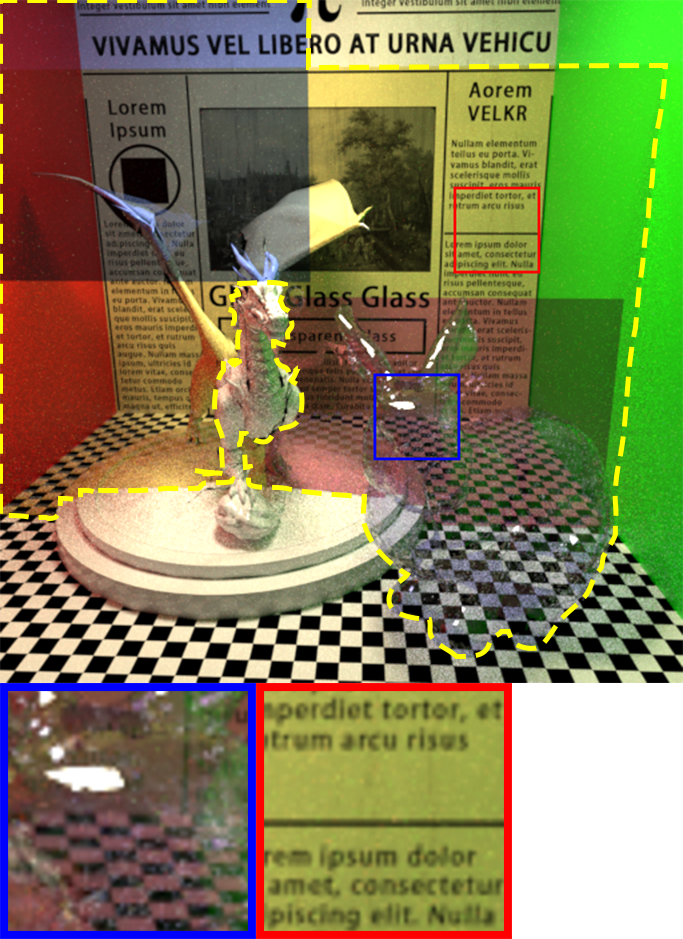}  & \compimage{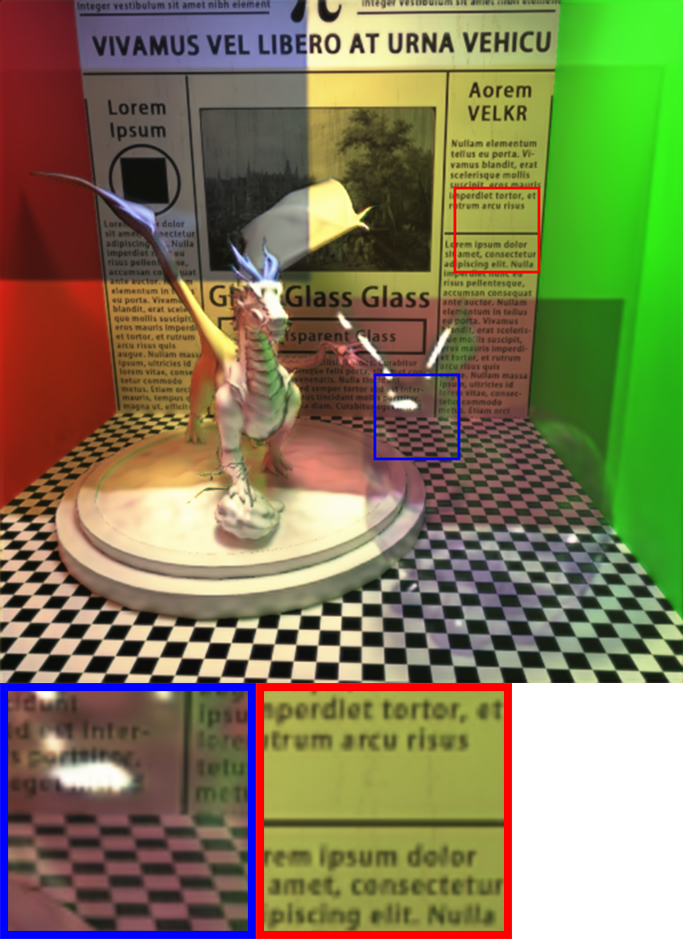} & \compimage{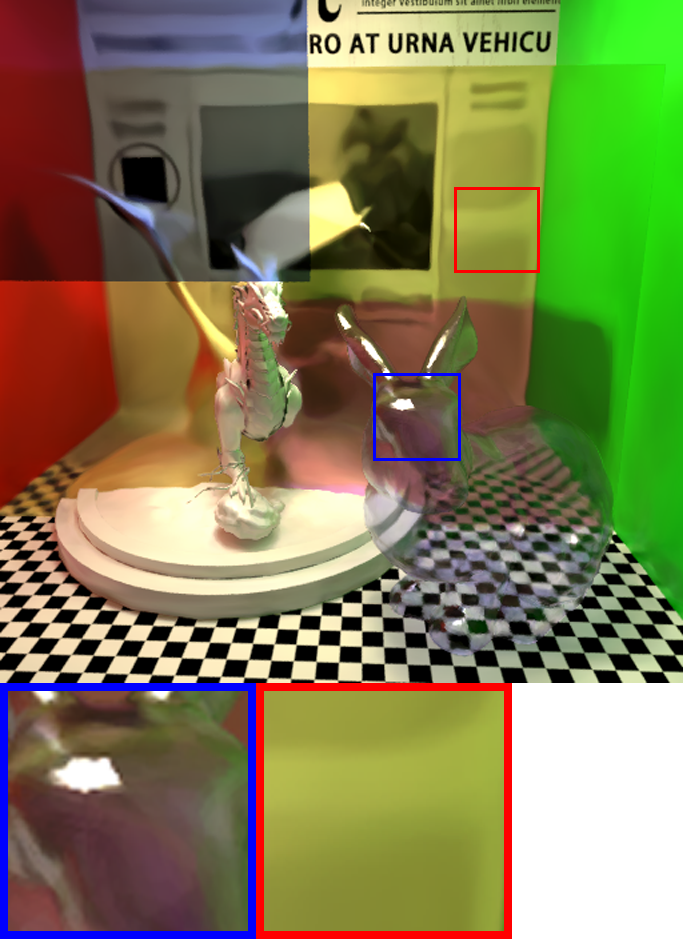} \\
\end{tabular}
\caption{\textbf{Qualitative comparison.} We compare with Active
Exploration in several challenging scenes. Note that
our method can preserve details of objects behind transparent surfaces with the
synthesis of global illumination. All scenes are rendered on test sets with
different camera views from the training set. The transparent areas are indicated by yellow dashed lines in the column of ground truths.}\label{fig:CompResult}
\end{figure*}
\renewcommand{\arraystretch}{1.0}
\newcommand{\comparebold}[2]{%
  \ifdim#1pt<#2pt
    \textbf{#1}%
  \else
    #1%
  \fi
}
\subsection{Experimental Settings}
\begin{table*}[t!]
  \centering
  \adjustbox{max width=1.0\textwidth}{
  \begin{tabular}{@{}rrccccccc@{}}
  \toprule
  \multirow{2}{*}{\small\textbf{Scene}} & \multirow{2}{*}{\textbf{Method}} & \multicolumn{6}{c}{\textbf{Metrics}} \\
  \cmidrule(lr){3-8}
  & & MAE $\downarrow$ & LIPIPS $\downarrow$ & DSSIM $\downarrow$ & PSNR $\uparrow$ & T.MAE $\downarrow$ & T.PSNR $\uparrow$ \\
\midrule
\multirow{2}{*}{BATHROOM}
& Active Exp. & 0.01475 & 0.02995 & 0.04840 & 30.85680 & 0.06170 & 23.72751 \\
& Neural G.I. & 0.00934 & 0.00961 & 0.02280 & 35.03145 & 0.03888 & 28.31231 \\
& Naive & 0.02131 & 0.04409 & 0.05622 & 29.17893 & 0.08208 & 22.41524 \\
& Ours & \textbf{0.00808} & \textbf{0.00686} & \textbf{0.01887} & \textbf{36.81351} & \textbf{0.02738} & \textbf{32.46141} \\
\midrule
\multirow{2}{*}{CORNELLBOX}
& Active Exp. & 0.03237 & 0.18584 & 0.20583 & 25.21868 & 0.10491 & 19.78125 \\
& Neural G.I. & 0.04093 & 0.12171 & 0.19213 & 24.13943 & 0.13141 & 17.84513 \\
& Naive & 0.04988 & 0.20916 & 0.23296 & 22.90424 & 0.15471 & 17.80871 \\
& Ours & \textbf{0.02142} & \textbf{0.01595} & \textbf{0.05436} & \textbf{29.75453} & \textbf{0.06087} & \textbf{25.44725} \\
\midrule
\multirow{2}{*}{DESK}
& Active Exp. & 0.01739 & 0.05523 & 0.06117 & 27.87549 & 0.15792 & 16.31566 \\
& Neural G.I. & 0.01733 & 0.04327 & 0.04985 & 29.34936 & 0.12360 & 17.96721 \\
& Naive & 0.03274 & 0.07845 & 0.07987 & 26.16205 & 0.19205 & 15.75951 \\
& Ours & \textbf{0.01658} & \textbf{0.01150} & \textbf{0.01782} & \textbf{34.82217} & \textbf{0.05487} & \textbf{28.24681} \\
\midrule
\multirow{2}{*}{CORNELLBOX8}
& Active Exp. & 0.02911 & 0.17719 & 0.20583 & 26.16428 & 0.08924 & 21.14504 \\
& Neural G.I. & 0.04307 & 0.19579 & 0.27657 & 22.83434 & 0.13061 & 17.90722 \\
& Naive & 0.05093 & 0.28167 & 0.31939 & 21.84069 & 0.15147 & 17.09835 \\
& Ours & \textbf{0.02064} & \textbf{0.04152} & \textbf{0.07311} & \textbf{28.56370} & \textbf{0.05721} & \textbf{24.25974} \\

  \bottomrule
  \end{tabular}
  }

  \caption{\textbf{Quantitative comparison.} We compare our approach with the naive baseline,
  Active Exploration and Neural Global Illumination on several scenes with diverse
  metrics. T.MAE and T.PSNR denote the MAE loss and PSNR on the areas of the image with transparent
  objects. Best result is denoted in \textbf{bold}.}
  \label{tab:loss_comparison}

  \end{table*}


The input and output resolution used for training are both $256\times256$ px. The super-sampling network uses $512\times512$ textures to super-sample the $256\times256$ \textit{GlassNet} rendered results to $512\times512$ px, improving the visual quality.
In addition, all inputs are positional encoded using the method proposed
by Mildenhall \etal. All types of inputs and ground
truth images are shown in Figure~\ref{fig:Inputs}. We used a relatively low
number of path tracing samples and did not pre-process the data with a denoiser.

We used Adam optimizer~\cite{kingma2014adam} to perform gradient descent
with different hyperparameters on each scene. The dataset details and hyperparameters
such as train/validation/test splits, learning rate and regularization for each
scene can be found in Appendix A. The training time varies from 5 to
10 hours using 8 RTX 2080 Ti GPUs, depending on the scene.

As evaluation metrics, we use Mean Absolute Error (MAE),
LPIPS~\cite{zhang2018unreasonable}, DSSIM~\cite{loza2006structural}, and Peak
Signal-to-Noise Ratio (PSNR). For transparency quality evaluation, we use PSNR
and MAE computed on pixels rendered with transparency (T.MAE and T.PSNR).

\subsection{Comparison with the State-of-the-Art}

\begin{figure}[tb]
  \centering
  \subfigure[Ground truth]{
    \includegraphics[width=0.148\textwidth]{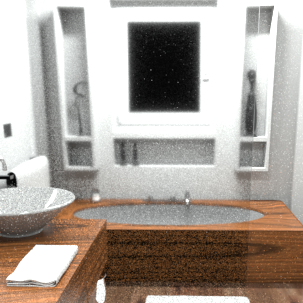}
  }
  \hfill
  \subfigure[Ours]{
    \includegraphics[width=0.148\textwidth]{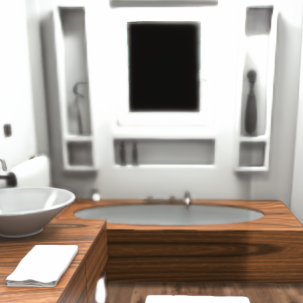}
  }
  \hfill
  \subfigure[Denoised Optix]{
    \includegraphics[width=0.148\textwidth]{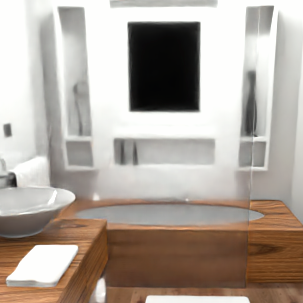}
  }
  \caption{\textbf{Comparison to real-time ray tracing.} The real-time path tracing denoiser failed to reconstruct the high frequency texture and small objects behind the glass.}
  \label{fig:RTRT}
\end{figure}

We compared with Active Exploration and Neural Global Illumination, as they are
the closest recent work. Comparisons were
carried out in 4 challenging scenes: BATHROOM, CORNELLBOX, DESK, and CORNELLBOX8 as shown in \cref{fig:ds}.
The BATHROOM was modified from the Bitterli
dataset~\cite{resources16}, while DESK and CORNELLBOX were creating using existing
models~\cite{Texturemontage05}, to focus on transparency effects. In CORNELLBOX8, We added 4 more transparent objects to CORNELLBOX to explore the performance under the condition of high number of transparent objects.
Qualitative results are shown in \cref{fig:teaser} and \cref{fig:CompResult}, while quantitative results are shown in
\cref{tab:loss_comparison}. A comparison with real-time path tracing method is also demonstrated in \cref{fig:RTRT}

From the qualitative comparison in \cref{fig:CompResult} we can see that
although Active Exploration performed well in many areas, without transparency
buffers, the neural network could not synthesize the area behind the
semi-transparent surfaces without losing fundamental details. In the DESK and CORNELLBOX
scenes, the text texture was almost impossible to render without provided texture
information. Instead, our method could synthesize all the texture accurately.
Furthermore, in the BATHROOM scene, all the detailed shapes behind the glass such as
taps were missing in the Active Exploration, where our method preserved them
well. The
quantitative comparison in \cref{tab:loss_comparison} corroborates the
visual improvement of our method, which achieves better values on most metrics,
with more significant improvements on parts of the image that need
transparency. In addition, the rendering results and comparison video are available in Video S1 in the Electronic Supplementary Material.

Overall, our method preserves the details of texture and geometry of shape
well. The global illumination effects such
as mirror reflection are also well synthesized. Our approach shows strong performance on preserve all the high
frequency texture, which is also shown on the wood texture of the BATHROOM
scene. Such details are extremely hard to be synthesized behind nearly
transparent surfaces without transparency buffers.

Our experiments show that the transparency buffer is essential for neural
network to render objects behind transparent surfaces accurately. Although the
global illumination effects are not the main focus of our method, we also
demonstrate good quality on indirect diffuse lighting, mirror reflection, and
soft shadows on par with existing approaches.

\begin{table}
  \centering
  \setlength\tabcolsep{4pt}
  { 

    \adjustbox{max width=1.0\textwidth}{
    \begin{tabular}{@{}rrcccc@{}}
      \toprule
      \multirow{2}{*}{\small\textbf{Scene}} & \multirow{2}{*}{\textbf{Method}} & \multicolumn{3}{c}{\textbf{Metrics}} \\
      \cmidrule(lr){3-5}
      & & DSSIM $\downarrow$ & T.MAE $\downarrow$ & T.PSNR $\uparrow$ \\
      \midrule
      \multirow{2}{*}{CORNELLBOX}
      & CNSR & 0.1502 & 0.0973 & 20.4854 \\
      & Ours & \textbf{0.0451} & \textbf{0.0457} & \textbf{26.2100} \\
      \midrule
      \multirow{2}{*}{DESK}
      & CNSR & 0.0539 & 0.1115 & 18.5453 \\
      & Ours & \textbf{0.0174} & \textbf{0.0533} & \textbf{28.6498} \\
      \bottomrule
    \end{tabular}
    }
    \caption{\textbf{Quantitative comparison with CNSR.} Best result is highlighted in \textbf{bold}.}
    \label{table:cnsr_compare}
  } 
\end{table}

\begin{figure*}[t!]
  \centering
  \begin{tabular}{cccc}
   & Ground Truth & Ours & CNSR \\
  \rotatebox[origin=c]{90}{CORNELLBOX}  & \compimage{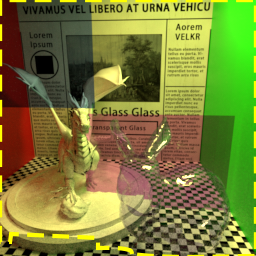}  & \compimage{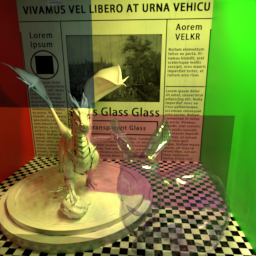} & \compimage{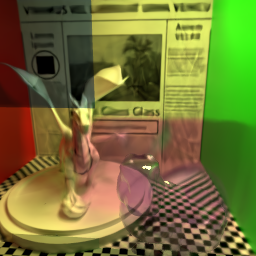} \\
  \end{tabular}
  \caption{\textbf{Rendering results of ours and Granskog \etal} Our method preserves the transparency quality better. The transparent areas are indicated by yellow dashed lines.}\label{fig:CNSRresult}
\end{figure*}

  We also compared our method to CNSR to further verify the effectiveness of our approach to transparency. We generated the dataset in the format CNSR requires and used the identical dataset to train the CNSR model and our model. The results are shown in \cref{fig:CNSRresult} and \cref{table:cnsr_compare}. The transparency quality rendered by our approach is significantly better. Note that this comparison experiment is only supplementary since the main focus of the two methods is different.

  In addition, we compared our approach to the real-time path tracing method in \cref{fig:RTRT}. We used 32 samples per pixel to generate the path tracing image with a standard Optix denoiser~\cite{chaitanya2017interactive}. The Optix denoiser failed to reconstruct the wood texture and the tape behind the glass, while our algorithm rendered such objects accurately. Due to the glass, the denoiser can no longer rely on albedo and normal AOV to gain extra information, causing the loss of details.
\begin{table}[ht]
  \centering
  \adjustbox{max width=0.6\textwidth}{
    \begin{tabular}{@{}rrccccc@{}}
      \toprule
      \multirow{2}{*}{\small\textbf{Size}} & \multirow{2}{*}{\textbf{Type}} & \multicolumn{4}{c}{\textbf{\# of Trans. Obj.}} \\
      \cmidrule(lr){3-6}
      & & 1 & 2 & 4 & 8 \\
      \midrule
      \multirow{2}{*}{$256\times256$}
         & Framerate (FPS)& {100} & {77} & {63} & {45}  \\
         & Frametime (ms) & {10} & {13} & {16} & {22}  \\
      \midrule
      \multirow{2}{*}{$512\times512$}
         & Framerate (FPS) & {50} & {37} & {32} & {23} \\
         & Frametime (ms)& {20} & {27} & {31} & {43} \\
         \midrule
      \multirow{2}{*}{$1024\times1024_{\text{CARN}}^\dagger$}
          & Framerate (FPS) & {24} & {21} & {20} & {18} \\
          & Frametime (ms)& {41} & {47} & {50} & {55} \\
      \midrule
      \multirow{2}{*}{$1024\times1024_{\text{NSRR}}^\dagger$}
          & Framerate (FPS) & {17} & {16} & {15} & {12} \\
          & Frametime (ms)& {56} & {62} & {66} & {78} \\
      \bottomrule
    \end{tabular}
  }

  \caption{\textbf{Performance evaluation.} We can interactively render all test cases. $\dagger$ $1024\times1024$ achieves by an external supersampling network, CARN or NSRR with temporal stability, as a post-processing procedure.}
  \label{tab:performance}

\end{table}

\begin{figure}
  \centering     
  \subfigure[PixelGenerator]{\includegraphics[width=0.225\textwidth]{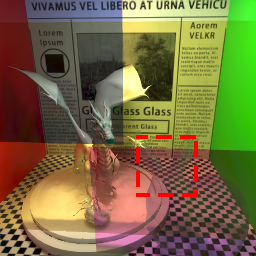}}
  \subfigure[CNN]{\includegraphics[width=0.225\textwidth]{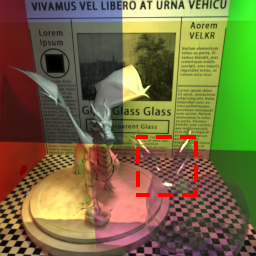}}
  \label{fig:ablation_pg}
  \caption{\textbf{PixelGenerator v.s. CNN.} Under the same training iterations, PixelGenerator completely failed to capture the glossy effects on the glass bunny.}
\end{figure}

\subsection{Ablation Study}

\begin{table}[t]
  \centering
  \begin{tabular}{rcccc}
  \toprule
             & MAE$\downarrow$  & LIPIPS$\downarrow$   & DSSIM$\downarrow$ & PSNR$\uparrow$    \\
  \midrule
  w/o P.E.   & 0.023711 & 0.026580 & 0.059718 & 28.937529  \\
  L1 Loss       & 0.023923 & 0.033820 & 0.072169 & 25.218679 \\
  Ours  & \textbf{0.021424} & \textbf{0.015953} & \textbf{0.054358} & \textbf{29.754528}  \\
  \bottomrule
  \end{tabular}
  \caption{\textbf{Impact of positional encoding and loss functions.} Our experiments show
  the combination of positional encoding and L1SSIM loss function
  achieves the best performance of the network. Best result is highlighted
  in \textbf{bold}.}\label{table:albation}
\end{table}

\begin{table}[t]
  \centering
  \begin{tabular}{rccc}
  \toprule
             & MAE  & MAE (inv. order)   & Difference    \\
  \midrule
  BATHROOM   & 0.00805 & 0.00805 & 0.00000  \\
  DESK       & 0.01658 & 0.01658 & 0.00000 \\
  CORNELLBOX  & 0.02125 & 0.02125 & 0.00000  \\
  \bottomrule
  \end{tabular}
  \caption{\textbf{MAE on different transparency buffer order.} We invert the order of transparency buffer, and achieve identical rendering results.}
  \label{table:OrderLoss}
\end{table}

Although various previous works such as Active Exploration and CNSR reported that the PixelGenerator has better
up-sampling performance than
CNN, we found that
PixelGenerator is not able to effectively synthesize the transparent structure
given the same amount of training time as CNN. We implemented the same
rendering network structure with PixelGenerator as the backbone network.
\Cref{fig:ablation_pg} shows that PixelGenerator failed to reconstruct the
glossy effects on transparent objects given $400$ training iterations.

\begin{figure}[tb]
  \centering
  \subfigure[Ground truth]{
    \includegraphics[width=0.148\textwidth]{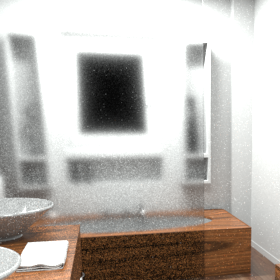}
  }
  \hfill
  \subfigure[Supersampling]{
    \includegraphics[width=0.148\textwidth]{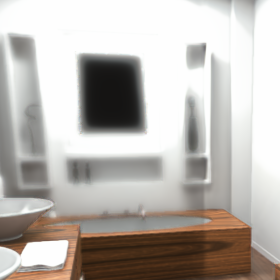}
  }
  \hfill
  \subfigure[Fine-tuned]{
    \includegraphics[width=0.148\textwidth]{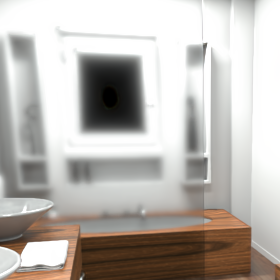}
  }
  \caption{\textbf{Fine-tuning v.s. Supersampling.} Fine-tuning network on
   $512\times512$ data has bad quality of roughness effects. The window frame
   should be  blurred out as in the ground truth. }
  \label{fig:albation_retrain}
\end{figure}

To overcome the bad up-sampling performance of CNN, an up-sampling network is
solely trained to super-sample the rendered images at two times the resolution.
Our experiment shows fine-tuning the model with a smaller high-resolution dataset
cannot achieve good visual effects. \Cref{fig:albation_retrain} shows that the
image generated by our up-sampling network has better visual quality,
especially on high-roughness transparent objects, than retraining the CNN.

We studied the influence of positional encoding on our method. Mildenhall
\etal show that positional encoding can help neural
networks learn high-frequency information; therefore, all the inputs to our
neural network are positional encoded. We also demonstrate that using the
combination of MAE loss and DSSIM loss can achieve the best quality, with quantitative results are shown in \cref{table:albation}. Furthermore, we verified
the effectiveness of our permutation-invariant algorithm.
\Cref{table:OrderLoss} shows that shuffled transparency buffer causes on effect
on our model performance.

\subsection{Performance Evaluation}

\begin{figure}[t]
  \centering
    \includegraphics[width=0.475\textwidth]{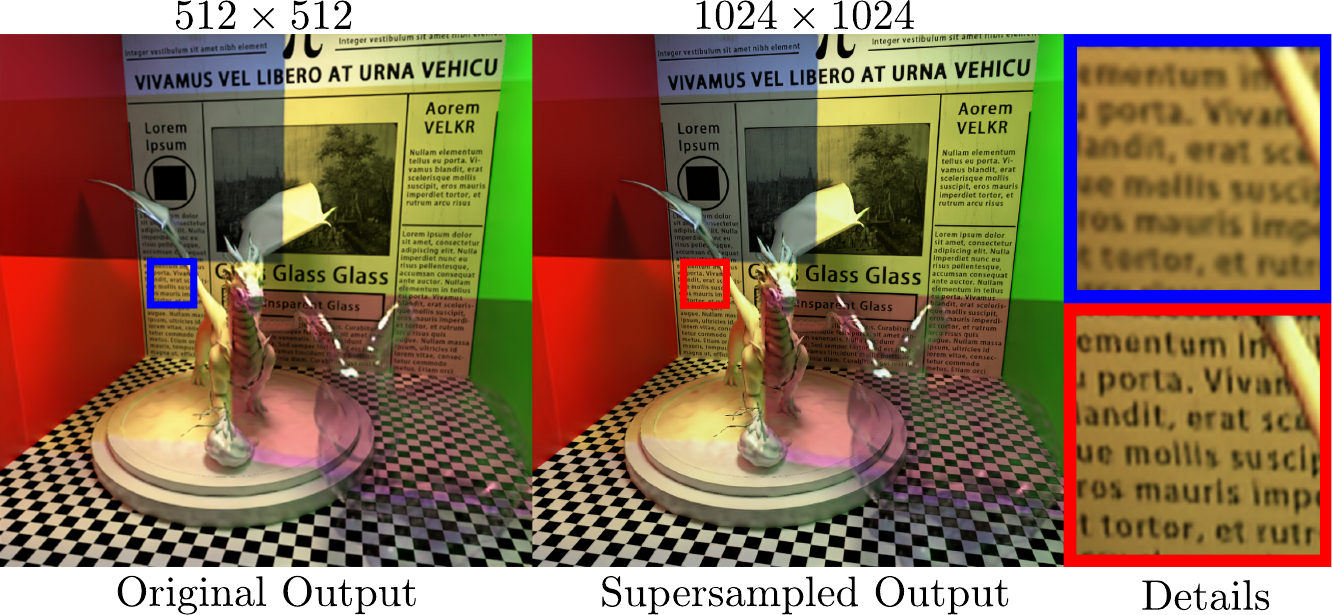}

    \caption{\textbf{Result in combining with other supersampling models. } We demonstrate that using external supersampling models such as NSRR can further supersample our results to $1024\times1024$.}
    \label{fig:carn_ss}
\end{figure}

\subsubsection{Real-time Performance}
\label{sec:Performance}
We tested our renderer on one RTX 4090, and the performance is shown in \cref{tab:performance}. The computation time of our algorithm does not significantly grow w.r.t. the complexity of meshes because the transparent buffers only separately record each transparent object instead of each primitive. The CORNELLBOX and DESK (30,344 and 38,400 triangles of transparent objects) show that our method can efficiently render complex meshes.

Furthermore, the neural blending algorithm shown in \cref{eq:SymmetricNN} also grows linearly w.r.t. the number of transparent objects given a specific resolution because $h$ is n $\mathcal{O}(1)$ function under specific input size. With the resolution of $512\times512$, one extra transparent object costs $3.3$ ms more time on average. We believe this cost can be further optimized by pruning the model.

Our approach can be further supersampled by other existing fast supersampling networks. As examples, we demonstrate that by using CARN\cite{ahn2018fast} and NSRR\cite{xiao2020neural} as post-processing models, our method can reach $1024\times1024$ resolution with anti-aliasing effects. The results can be found in \cref{tab:performance}, \cref{fig:carn_ss}, and Video S1 in the Electronic Supplementary Material.

\begin{figure}
  \centering
  \begin{tikzpicture}
    \begin{axis}[
        xlabel=Number of transparent objects,
        ylabel=Required memory (Megabytes),
        xmin=0, xmax=9,
        ymin=0, ymax=38,
        xtick={1,2,4,8},
        ytick={2, 4, 6, 8, 10, 12, 14, 16, 18, 20, 22, 24, 26, 28, 30, 32, 34, 36},
        legend pos=north west,
        grid=both,
        grid style={line width=.1pt, draw=gray!10},
        major grid style={line width=.2pt,draw=gray!50},
        legend style={font=\small} 
        ]
    \addplot[mark=*,blue] plot coordinates {
        (1,4.456)
        (2,8.912)
        (4,8.912)
        (8,8.912)
    };
    \addlegendentry{Ours}
  
    \addplot[color=red,mark=x]
        plot coordinates {
          (1,4.456)
          (2,8.912)
          (4,17.824)
          (8,35.648)
        };
    \addlegendentry{Traditional}
    \end{axis}
  \end{tikzpicture}
\caption{\textbf{Buffer memory usage comparison.} Under $256 \times 256$, one transparent object buffer causes around 4.5 megabytes of memory. Our blending algorithm is summation based and thus do not require space to store each transparency buffer.}
\label{fig:MemUsage}
\end{figure}
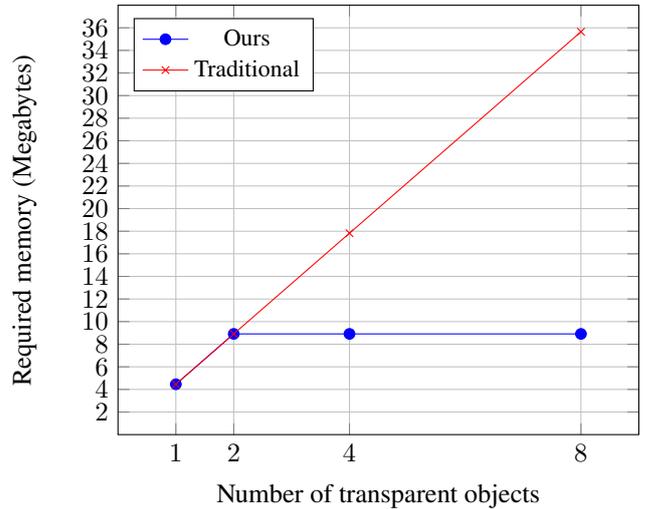

\subsubsection{Memory Efficiency}
The permutation invariant blending algorithm also has a memory efficiency
advantage over existing approaches. With traditional methods, the model has to receive all transparency
buffers in order to make inference, so that the memory costs grow linearly
w.r.t. the number of transparent objects. However, with our proposed algorithm,
there is no additional framebuffer costs because the model can operate on a
single buffer of each transparent objects, and then add it to the summed latent
representation. The memory usage comparison is plotted in \cref{fig:MemUsage}
approximated by framebuffer usage. We can see that our permutation
invariant blending algorithm has constant space growth w.r.t. the number of
transparent objects.

\begin{table}[t]
  \centering
  \begin{tabular}{rcc}
    
  \toprule
             & RAM (GB)  & VRAM (GB)  \\
  \midrule
  Active Exploration   & 2.50 & 11  \\
  Ours       & \textbf{0.68} & \textbf{4} \\
  \bottomrule
  \end{tabular}
  \caption{\textbf{Runtime memory usage comparison.} We compare our method with Active Exploration on $512 \times 512$ with three transparent objects. Least memory usage is highlighted
  in \textbf{bold}.}\label{tab:RMemUsage}
\end{table}

We also compare the actual runtime RAM and VRAM usage of our method with that of Active Exploration. The experiment uses the DESK scene with the resolution of $512\times512$ to test the memory performance. \Cref{tab:RMemUsage} shows our implementation significantly reduces the required runtime memory.

\section{Limitations and Discussion}

\begin{figure}[tb]
  \centering
  \subfigure[Ground truth]{
    \includegraphics[width=0.225\textwidth]{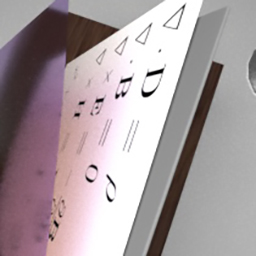}
  }
  \hfill
  \subfigure[Ours]{
    \includegraphics[width=0.225\textwidth]{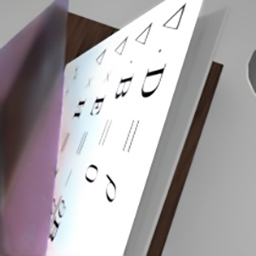}
  }
  \caption{\textbf{Failed case.} When the refraction effects dominate, our
  method can no longer accurately render surfaces under transparent objects
  because G-buffer cannot provide useful information.}
  \label{fig:limitation}
\end{figure}

Although our method shows significant improvements in neural rendering of
scenes with transparent objects with full OIT,
our method relies on two fundamental assumptions about scenes: the presence of
transparent surfaces with a low refraction index and the absence of
participating media. However, when refraction occurs, the shading information
provided by the G-buffer is no longer accurate, which becomes an even more
significant issue on spherical surfaces. This limitation is illustrated in
\cref{fig:limitation}. Our rendering network currently struggles to produce
correct results when strong refraction effects are present.  Additionally, due
to the use of G-buffers, our framework is unable to represent participating
media, leading to the ineffective synthesis of volumetric effects.

As a problem of neural rendering in general, current neural rendering approaches perform relatively poorly on high-resolution images. Although our method uses CARN or NSRR as a post-processing supersampling procedure to boost the resolution further to $1024 \times 1024$ as described in \cref{sec:Performance}, further attempts to raise resolution will significantly affect the performance, causing low framerates. Further research may focus on the performance of neural rendering in high-resolution settings.

Another potential direction for future research is incorporating path prediction
techniques, such as the method proposed by Li \etal~\cite{li2020through}, into
scene representation. Furthermore, exploring efficient approaches for
representing participating media on both the input and neural network sides
presents an intriguing research topic.

Several orthogonal methods can be combined with our work. For example, the
Monte Carlo sampling technique proposed in Active
Exploration can be utilized to improve the effects
such as perfect mirror reflection and caustic effects. Neural Global Illumination uses radiance cues as inputs to the model, which improves the scene representation learning.

\appendix

\subsection*{Appendix}

\Cref*{table:hyper} records the key scene settings and hyperparameters during the dataset generation, training, and evaluation.
\begin{table*}[t!]
  \centering
  \begin{tabular}{lcccccccc}
  \toprule
  \textbf{Scene} & \textbf{Train} & \textbf{Validation} & \textbf{Test} & \textbf{S.P.P.} & \textbf{L.R.} & \textbf{W.D.} & \textbf{\# of Trans. Obj.} & \textbf{Variable} \\
  \midrule
  BATHROOM    & 1800 & 200 & 128 & 8192 & 0.0001 & 0.00075 & 1 & 1 (Roughness) \\
  CORNELLBOX  & 1800 & 200 & 128 & 4096 & 0.0001 & 0.0001  & 4 & 4 (Position and Lighting) \\
  DESK        & 1800 & 200 & 128 & 4096 & 0.0001 & 0.0001  & 3 & 3 (Color) \\
  CORNELLBOX8 & 900  & 100 & 128 & 4096 & 0.0001 & 0.0001  & 8 & 7 (Position and Lighting) \\
  \bottomrule
  \end{tabular}
  \caption{\textbf{Scene settings and hyperparameter choices.} S.P.P. indicates the number of samples per pixel using during the ground truth generation. L.R. refers to learning rate and W.D. refers to weight decay.}
  \label{table:hyper}
\end{table*}

\bibliographystyle{IEEEtran}
\bibliography{main.bib}

\begin{thebibliography}{10}
\providecommand{\url}[1]{#1}
\csname url@samestyle\endcsname
\providecommand{\newblock}{\relax}
\providecommand{\bibinfo}[2]{#2}
\providecommand{\BIBentrySTDinterwordspacing}{\spaceskip=0pt\relax}
\providecommand{\BIBentryALTinterwordstretchfactor}{4}
\providecommand{\BIBentryALTinterwordspacing}{\spaceskip=\fontdimen2\font plus
\BIBentryALTinterwordstretchfactor\fontdimen3\font minus \fontdimen4\font\relax}
\providecommand{\BIBforeignlanguage}[2]{{%
\expandafter\ifx\csname l@#1\endcsname\relax
\typeout{** WARNING: IEEEtran.bst: No hyphenation pattern has been}%
\typeout{** loaded for the language `#1'. Using the pattern for}%
\typeout{** the default language instead.}%
\else
\language=\csname l@#1\endcsname
\fi
#2}}
\providecommand{\BIBdecl}{\relax}
\BIBdecl

\bibitem{diolatzis2022active}
S.~Diolatzis, J.~Philip, and G.~Drettakis, ``Active exploration for neural global illumination of variable scenes,'' \emph{ACM Transactions on Graphics (TOG)}, vol.~41, no.~5, pp. 1--18, 2022.

\bibitem{kajiya1986rendering}
J.~T. Kajiya, ``The rendering equation,'' in \emph{Proceedings of the 13th annual conference on Computer graphics and interactive techniques}, 1986, pp. 143--150.

\bibitem{lafortune1996rendering}
E.~P. Lafortune and Y.~D. Willems, ``Rendering participating media with bidirectional path tracing,'' in \emph{Rendering Techniques '96}, X.~Pueyo and P.~Schr{\"o}der, Eds., 1996, pp. 91--100.

\bibitem{greger1998irradiance}
G.~Greger, P.~Shirley, P.~M. Hubbard, and D.~P. Greenberg, ``The irradiance volume,'' \emph{IEEE Computer Graphics and Applications}, vol.~18, no.~2, pp. 32--43, 1998.

\bibitem{ren2013global}
P.~Ren, J.~Wang, M.~Gong, S.~Lin, X.~Tong, and B.~Guo, ``Global illumination with radiance regression functions.'' \emph{ACM Trans. Graph.}, vol.~32, no.~4, pp. 130--1, 2013.

\bibitem{eslami2018neural}
S.~A. Eslami, D.~Jimenez~Rezende, F.~Besse, F.~Viola, A.~S. Morcos, M.~Garnelo, A.~Ruderman, A.~A. Rusu, I.~Danihelka, K.~Gregor \emph{et~al.}, ``Neural scene representation and rendering,'' \emph{Science}, vol. 360, no. 6394, pp. 1204--1210, 2018.

\bibitem{granskog2020compositional}
J.~Granskog, F.~Rousselle, M.~Papas, and J.~Nov{\'a}k, ``Compositional neural scene representations for shading inference,'' \emph{ACM Transactions on Graphics (TOG)}, vol.~39, no.~4, pp. 135--1, 2020.

\bibitem{rainer2022neural}
\BIBentryALTinterwordspacing
G.~Rainer, A.~Bousseau, T.~Ritschel, and G.~Drettakis, ``Neural precomputed radiance transfer,'' \emph{Computer Graphics Forum (Proceedings of the Eurographics conference)}, vol.~41, no.~2, April 2022. [Online]. Available: \url{http://www-sop.inria.fr/reves/Basilic/2022/RBRD22}
\BIBentrySTDinterwordspacing

\bibitem{gao2022neural}
D.~Gao, H.~Mu, and K.~Xu, ``Neural global illumination: Interactive indirect illumination prediction under dynamic area lights,'' \emph{IEEE Transactions on Visualization and Computer Graphics}, 2022.

\bibitem{porter1984compositing}
T.~Porter and T.~Duff, ``Compositing digital images,'' in \emph{Proceedings of the 11th annual conference on Computer graphics and interactive techniques}, 1984, pp. 253--259.

\bibitem{carpenter1984buffer}
L.~Carpenter, ``The a-buffer, an antialiased hidden surface method,'' in \emph{Proceedings of the 11th annual conference on Computer graphics and interactive techniques}, 1984, pp. 103--108.

\bibitem{jouppi1999z}
N.~P. Jouppi and C.-F. Chang, ``Z 3: an economical hardware technique for high-quality antialiasing and transparency,'' in \emph{Proceedings of the ACM SIGGRAPH/EUROGRAPHICS workshop on Graphics hardware}, 1999, pp. 85--93.

\bibitem{everitt2001interactive}
C.~Everitt, ``Interactive order-independent transparency,'' \emph{White paper, nVIDIA}, vol.~2, no.~6, p.~7, 2001.

\bibitem{saito1990comprehensible}
T.~Saito and T.~Takahashi, ``Comprehensible rendering of 3-d shapes,'' in \emph{Proceedings of the 17th annual conference on Computer graphics and interactive techniques}, 1990, pp. 197--206.

\bibitem{pangerl2009deferred}
D.~Pangerl, ``Deferred rendering transparency,'' \emph{ShaderX7: Advanced Rendering Techniques, ShaderX series}, pp. 217--224, 2009.

\bibitem{mara2013lighting}
M.~Mara, M.~McGuire, and D.~Luebke, ``Lighting deep g-buffers: Single-pass, layered depth images with minimum separation applied to indirect illumination,'' \emph{NVIDIA Corporation}, 2013.

\bibitem{xin2020lightweight}
H.~Xin, S.~Zheng, K.~Xu, and L.-Q. Yan, ``Lightweight bilateral convolutional neural networks for interactive single-bounce diffuse indirect illumination,'' \emph{IEEE Transactions on Visualization and Computer Graphics}, vol.~28, no.~4, pp. 1824--1834, 2020.

\bibitem{keller2015path}
\BIBentryALTinterwordspacing
A.~Keller, L.~Fascione, M.~Fajardo, I.~Georgiev, P.~Christensen, J.~Hanika, C.~Eisenacher, and G.~Nichols, ``The path tracing revolution in the movie industry,'' in \emph{ACM SIGGRAPH 2015 Courses}, 2015, pp. 1--7. [Online]. Available: \url{https://doi.org/10.1145/2776880.2792699}
\BIBentrySTDinterwordspacing

\bibitem{ritschel2012state}
T.~Ritschel, C.~Dachsbacher, T.~Grosch, and J.~Kautz, ``The state of the art in interactive global illumination,'' in \emph{Computer graphics forum}, vol.~31, 2012, pp. 160--188.

\bibitem{o2018precomputed}
Y.~O'Donnell, ``Precomputed global illumination in frostbite,'' 2018.

\bibitem{cohen1993radiosity}
M.~F. Cohen, J.~R. Wallace, and P.~Hanrahan, ``Radiosity and realistic image synthesis,'' 1993.

\bibitem{mcguire2017real}
M.~McGuire, M.~Mara, D.~Nowrouzezahrai, and D.~Luebke, ``Real-time global illumination using precomputed light field probes,'' in \emph{Proceedings of the 21st ACM SIGGRAPH symposium on interactive 3D graphics and games}, 2017, pp. 1--11.

\bibitem{tewari2020state}
A.~Tewari, O.~Fried, J.~Thies, V.~Sitzmann, S.~Lombardi, K.~Sunkavalli, R.~Martin-Brualla, T.~Simon, J.~Saragih, M.~Nie{\ss}ner \emph{et~al.}, ``State of the art on neural rendering,'' in \emph{Computer Graphics Forum}, vol.~39, 2020, pp. 701--727.

\bibitem{raghavan2023neural}
N.~Raghavan, Y.~Xiao, K.-E. Lin, T.~Sun, S.~Bi, Z.~Xu, T.-M. Li, and R.~Ramamoorthi, ``Neural free-viewpoint relighting for glossy indirect illumination,'' in \emph{Computer Graphics Forum}, vol.~42, 2023, p. e14885.

\bibitem{qi2017pointnet}
C.~R. Qi, H.~Su, K.~Mo, and L.~J. Guibas, ``Pointnet: Deep learning on point sets for 3d classification and segmentation,'' in \emph{Proceedings of the IEEE conference on computer vision and pattern recognition}, 2017, pp. 652--660.

\bibitem{bloem2020probabilistic}
B.~Bloem-Reddy and Y.~W. Teh, ``Probabilistic symmetries and invariant neural networks,'' \emph{The Journal of Machine Learning Research}, vol.~21, no.~1, pp. 3535--3595, 2020.

\bibitem{heitz2016real}
E.~Heitz, J.~Dupuy, S.~Hill, and D.~Neubelt, ``Real-time polygonal-light shading with linearly transformed cosines,'' \emph{ACM Transactions on Graphics (TOG)}, vol.~35, no.~4, pp. 1--8, 2016.

\bibitem{wloka2003batch}
M.~Wloka, ``Batch, batch, batch: What does it really mean,'' 2003.

\bibitem{pantazopoulos2002occlusion}
I.~Pantazopoulos and S.~Tzafestas, ``Occlusion culling algorithms: A comprehensive survey,'' \emph{Journal of Intelligent and Robotic Systems}, vol.~35, pp. 123--156, 2002.

\bibitem{meshkin2007sort}
H.~Meshkin, ``Sort-independent alpha blending,'' \emph{GDC Talk}, vol.~2, no.~4, 2007.

\bibitem{salvi2014multi}
M.~Salvi and K.~Vaidyanathan, ``Multi-layer alpha blending,'' in \emph{Proceedings of the 18th meeting of the ACM SIGGRAPH Symposium on Interactive 3D Graphics and Games}, 2014, pp. 151--158.

\bibitem{isola2017image}
P.~Isola, J.-Y. Zhu, T.~Zhou, and A.~A. Efros, ``Image-to-image translation with conditional adversarial networks,'' in \emph{Proceedings of the IEEE conference on computer vision and pattern recognition}, 2017, pp. 1125--1134.

\bibitem{mildenhall2021nerf}
B.~Mildenhall, P.~P. Srinivasan, M.~Tancik, J.~T. Barron, R.~Ramamoorthi, and R.~Ng, ``Nerf: Representing scenes as neural radiance fields for view synthesis,'' \emph{Communications of the ACM}, vol.~65, no.~1, pp. 99--106, 2021.

\bibitem{Jakob2020DrJit}
W.~Jakob, S.~Speierer, N.~Roussel, and D.~Vicini, ``Dr.jit: A just-in-time compiler for differentiable rendering,'' \emph{Transactions on Graphics (Proceedings of SIGGRAPH)}, vol.~41, no.~4, Jul. 2022.

\bibitem{wang2004image}
Z.~Wang, A.~C. Bovik, H.~R. Sheikh, and E.~P. Simoncelli, ``Image quality assessment: from error visibility to structural similarity,'' \emph{IEEE transactions on image processing}, vol.~13, no.~4, pp. 600--612, 2004.

\bibitem{kingma2014adam}
D.~P. Kingma and J.~Ba, ``Adam: A method for stochastic optimization,'' \emph{arXiv preprint arXiv:1412.6980}, 2014.

\bibitem{zhang2018unreasonable}
R.~Zhang, P.~Isola, A.~A. Efros, E.~Shechtman, and O.~Wang, ``The unreasonable effectiveness of deep features as a perceptual metric,'' in \emph{Proceedings of the IEEE conference on computer vision and pattern recognition}, 2018, pp. 586--595.

\bibitem{loza2006structural}
A.~Loza, L.~Mihaylova, N.~Canagarajah, and D.~Bull, ``Structural similarity-based object tracking in video sequences,'' in \emph{2006 9th International Conference on Information Fusion}, 2006, pp. 1--6.

\bibitem{resources16}
B.~Bitterli, ``Rendering resources,'' 2016, https://benedikt-bitterli.me/resources/.

\bibitem{Texturemontage05}
K.~Zhou, X.~Wang, Y.~Tong, M.~Desbrun, B.~Guo, and H.-Y. Shum, ``Texturemontage: Seamless texturing of arbitrary surfaces from multiple images,'' \emph{ACM Transactions on Graphics}, vol.~24, no.~3, pp. 1148--1155, 2005.

\bibitem{chaitanya2017interactive}
C.~R.~A. Chaitanya, A.~S. Kaplanyan, C.~Schied, M.~Salvi, A.~Lefohn, D.~Nowrouzezahrai, and T.~Aila, ``Interactive reconstruction of monte carlo image sequences using a recurrent denoising autoencoder,'' \emph{ACM Transactions on Graphics (TOG)}, vol.~36, no.~4, pp. 1--12, 2017.

\bibitem{ahn2018fast}
N.~Ahn, B.~Kang, and K.-A. Sohn, ``Fast, accurate, and lightweight super-resolution with cascading residual network,'' in \emph{Proceedings of the European conference on computer vision (ECCV)}, 2018, pp. 252--268.

\bibitem{xiao2020neural}
L.~Xiao, S.~Nouri, M.~Chapman, A.~Fix, D.~Lanman, and A.~Kaplanyan, ``Neural supersampling for real-time rendering,'' \emph{ACM Transactions on Graphics (TOG)}, vol.~39, no.~4, pp. 142--1, 2020.

\bibitem{li2020through}
Z.~Li, Y.-Y. Yeh, and M.~Chandraker, ``Through the looking glass: Neural 3d reconstruction of transparent shapes,'' in \emph{Proceedings of the IEEE/CVF Conference on Computer Vision and Pattern Recognition}, 2020, pp. 1262--1271.

\end{thebibliography}

\end{document}